\PassOptionsToPackage{pdfpagelabels=false}{hyperref} 
\documentclass[usenatbib]{mnras_tex_edited}

\usepackage{newtxtext,newtxmath}

\usepackage[T1]{fontenc}
\usepackage{ae,aecompl}

\usepackage{graphicx}	%
\usepackage{pdfpages}
\usepackage{float}
\usepackage{booktabs}
\usepackage{multirow}
\usepackage{array}
\usepackage{soul}
\usepackage{xcolor}
\definecolor{darkolivegreen}{rgb}{0.33, 0.42, 0.18}

\usepackage{color}

\newcommand{\kms}{km/s}
\newcommand{\msun}{$M_{\odot}$}

\def\aj{\rm{AJ}}
\def\araa{\rm{ARA\&A}}
\def\apj{\rm{ApJ}}
\def\apjl{\rm{ApJ}}
\def\apjs{\rm{ApJS}}
\def\aap{\rm{A\&A}}

\def\mnras{\rm{MNRAS}}
\def\prd{\rm{PRD}}
\def\prl{\rm{PRL}}

\def\pasp{\rm{PASP}}

\def\nat{\rm{Nature}}

\def\cqg{\rm{CQG}}
\def\lrr{\rm{LRR}}

\title[MBH binary inspiral and spins]{Massive black hole binary inspiral and spin evolution in a cosmological framework}

\author[M. Sayeb et al.]{Mohammad Sayeb$^{1}$\thanks{E-mail: \href{mailto:sayebms1@ufl.edu}{sayebms1@ufl.edu}},
Laura Blecha$^{1}$,
Luke Zoltan Kelley$^{2}$,
Davide Gerosa$^{3}$,\and
Michael Kesden$^{4}$,
July Thomas$^{1}$
\bigskip \\
$^{1}$Department of Physics, University of Florida, Gainesville, FL 32611, USA\\
$^{2}$Center for Interdisciplinary Exploration and Research in Astrophysics (CIERA) and \\\hspace{0.7em}Department of Physics \& Astronomy, Northwestern University, Evanston, IL 60201, USA\\
$^{3}$School of Physics and Astronomy \&	 Institute for Gravitational Wave Astronomy, University of Birmingham, Birmingham, B15 2TT, UK\\
$^{4}$Department of Physics, University of Texas at Dallas, Richardson, TX 75080, USA\\
}

\pubyear{2020}

\begin{document}
\label{firstpage}
\pagerange{\pageref{firstpage}--\pageref{lastpage}}
\maketitle

\begin{abstract} 
Massive black hole (MBH) binary inspiral time scales are uncertain, and their spins are even
more poorly constrained. Spin misalignment introduces asymmetry in the gravitational radiation, which imparts a recoil kick to the merged MBH. Understanding how MBH binary spins evolve is crucial for determining their recoil velocities, their gravitational wave (GW) waveforms detectable with LISA, as well as their retention rate in galaxies. Here we introduce a sub-resolution model for gas- and GW-driven MBH binary spin evolution using accreting MBHs from the Illustris cosmological hydrodynamics simulations. We also model binary inspiral via dynamical friction, stellar scattering, viscous gas drag, and GW emission. Our model assumes that the circumbinary disk always removes angular momentum from the binary. It also assumes differential accretion, which causes greater alignment of the secondary MBH spin in unequal-mass mergers. We find that 47\% of the MBHs in our population merge by $z=0$. Of these, 19\% have misaligned primaries and 10\% have misaligned secondaries at the time of merger in our
fiducial model with initial eccentricity of 0.6 and accretion rates from Illustris. The MBH
misalignment fraction depends strongly on the accretion disc parameters, however. Reducing
accretion rates by a factor of 100, in a thicker disc, yields 79\% and 42\% misalignment for
primaries and secondaries, respectively. Even in the more conservative fiducial model, more
than 12\% of binaries experience recoils of > 500km/s, which could displace them at least
temporarily from galactic nuclei. We additionally find that a significant number of systems
experience strong precession.

    \hspace{1em}
\end{abstract}


\section{Introduction}

 Numerous studies have found a correlation between the masses of massive black holes (MBHs) and the 
stellar bulges of their host galaxies
 \citep[e.g.][]{2009ApJ...698..198G,2013ARA&A..51..511K}.
The origin of these unexpected correlations is still an open
question, but galaxy mergers are likely to play a role \citep{2015ARA&A..53...51S}.
A satellite galaxy can gravitationally influence the gas in its host galaxy, and significantly reduce 
its angular momentum, leading to its
in-fall towards the galactic center \citep{1992ApJ...393..484B,1992ApJ...400..460H}.
This can supply fuel to the MBH \citep{2005Natur.433..604D, 2005MNRAS.361..776S} and may also trigger 
a burst of star formation around the nucleus \citep{1996ARA&A..34..749S,2009ApJS..182..216K,2012MNRAS.421.1539N,
2013MNRAS.428.2529H}.

Galaxy mergers can also lead to the formation
of a bound MBH binary \citep{1980Natur.287..307B,1981A&A...104..218R}.
Interactions with stars and gas in the nucleus will shrink the binary's orbit
until general relativistic 
effects become
important. At this stage, the binary is driven to merger by gravitational wave (GW)
emission.

Crucially, the formation of a MBH binary does not always guarantee merging within a Hubble time. The binary will go through different
phases of evolution that can be categorized into four main stages \citep{1980Natur.287..307B}.
The inspiral is first driven by dark matter, stellar, and gas dynamical friction (DF).
At separations of $\sim$ a few parsec, when a bound binary forms, interactions with individual low-angular-momentum
stars become important. At this stage, the 
binary loses energy through individual stellar scatterings. Because the range of the available momenta that satisfy the 
requirement for stellar scattering represents a cone in phase space, this stage is typically referred to as 
loss-cone (LC) star scattering \citep{2013CQGra..30x4005M}.
The stars 
are scattered out of the system,  
which removes energy from the MBH binary and shrinks its separation to a few 
tenths of a pc \citep{2013CQGra..30x0301M}. In gas-rich systems, further shrinking of the binary
separation can happen through gas-driven inspiral where orbital energy and angular momentum 
are imparted to a circumbinary disc (CBD).  
Finally, energy loss through GW emission takes over and leads the binary to merger
In general, at any binary separation a combination of these mechanisms is at play  and 
 determines the merger timescale and fate of the MBHs \citep[cf.][]{2020MNRAS.498.2219V}.

MBH mergers in the lower mass range of $\rm M \lesssim 10^7$ \msun\
emit GWs at $\sim$mHz frequencies 
which can be detected by the future Laser Interferometer Space Antenna (LISA) 
\citep{2017arXiv170200786A}.
Very low frequency ($\sim$nHz) GWs emitted by $\rm M \gtrsim 10^8$ \msun\ MBH 
binaries are detectable by pulsar timing arrays \citep[PTAs;][]{1979ApJ...234.1100D,
1978SvA....22...36S,2012PhRvL.109h1104M,2015RPPh...78l4901L, 2015arXiv151107869B}.

Merging MBH binaries with unequal masses or spins produce asymmetric GW radiation, which in turn imparts
a recoil velocity to the remnant MBH \citep{1962PhRv..128.2471P,1983MNRAS.203.1049F,1973ApJ...183..657B}.  
Recoils can reduce merger 
rates \citep{2009CQGra..26i4033S} and affect the growth of MBHs 
and the co-evolution of the MBH-galaxy system \citep{2008MNRAS.383.1079V,
2008ApJ...678..780G,2008MNRAS.390.1311B,2011MNRAS.412.2154B, 2011MNRAS.414.3656S}.
Large recoil velocities of $\gtrsim1000$ \kms, produced by some simulations, can even escape massive elliptical galaxies \citep{2007ApJ...667L.133S, 2015MNRAS.446...38G}. 
 Ejected MBHs might be rare at low redshifts, but in the early
universe, with smaller escape speeds and larger merger rates, their
frequency might be higher 
\citep{2003ApJ...582..559V, 2004ApJ...604..484M, 2011ApJ...742...13B,2016MNRAS.456..961B} and could lead 
to a population of intergalactic MBHs \citep[e.g.][]{2008ApJ...678L..81K}. 
This is important for the
early phase of MBH growth from stellar-mass or intermediate-mass precursors and 
consequently for the frequency of GW signals and event rates detectable by LISA \citep{2007MNRAS.377.1711S}. 
It could also have important repercussions on the observed scatter in the MBH mass and bulge velocity dispersion
relations \citep{2006MNRAS.368.1381L,2008MNRAS.383.1079V, 2008ApJ...678..780G,2011MNRAS.412.2154B}.

Following a MBH recoil event, the most tightly bound stars and gas will remain with the MBH while the gas and stars at larger radii will 
be left behind \citep{2004ApJ...607L...9M, 2006MNRAS.367.1746M, 2004ApJ...606L..17M, 2007PhRvL..99d1103L}. This can create a 
relative redshift that can be observed as an offset between broad and narrow lines. One such GW recoil candidate identified was  SDSSJ092712.65+294344.0 ---an active galactic nucleus (AGN) with a 2650 \kms\ shift between its broad and narrow emission 
lines \citep{2008ApJ...678L..81K}. 
Further study showed that this effect could be caused by other astrophysical 
phenomena such as a sub-parsec binary \citep{2009ApJ...697..288B}, or a large and small galaxy interacting near the 
center of a rich cluster \citep{2009ApJ...695..363H}. 
CID-42 
is another promising candidate presenting both
spatial and spectroscopic offset signatures, but other interpretations are possible \citep{2010ApJ...717..209C,2012ApJ...752...49C,2013MNRAS.428.1341B}.  
A growing number of other GW recoil candidates have been identified \citep{2012AdAst2012E..14K},
but none have yet provided unambiguous 
evidence for a recoiling MBH \citep{2010ApJ...717L.122R, 2010ApJ...717..209C,
2010ApJ...717L...6B, 2014MNRAS.445..515K, 2017A&A...600A..57C}.
The anisotropic emission of linear momentum that causes recoils is imprinted in the emitted GW signals, thus making merger kicks a potential observable for GW interferometers \citep{2016PhRvL.117a1101G,2018PhRvL.121s1102C,
2020PhRvL.124j1104V}.

In gas-rich systems, a key element is the interaction of the MBHs with their accretion discs (i.e. the CBD phase). 
There have been extensive studies and simulations of the interactions of MBHs with the circumbinary disc as 
isolated systems 
\citep{1996ApJ...467L..77A,1998ApJ...506L..97N, 2002A&A...387..550G,2008ApJ...672...83M, 2009MNRAS.399.2249P,
2010ApJ...708..485H,2012ApJ...749..118S,2013MNRAS.436.2997D,2014ApJ...783..134F,
2015ApJ...807..131S, 2017ApJ...838...42B,2017ApJ...835..199R,2018MNRAS.476.2249T}.
The long-standing consensus on gas-rich systems says that higher accretion rates can lead to dynamical torques and viscous drag contributing significantly to shrinking of the binary separation \citep{1980Natur.287..307B,2000ApJ...532L..29G,2002ApJ...567L...9A,2005ApJ...630..152E, 2008ApJ...672...83M,2009ApJ...700.1952H,2009MNRAS.398.1392L, 2012A&A...545A.127R,2016ApJ...827..111R,2017MNRAS.469.4258T}. This 
effect is enhanced in galaxy mergers which drive more gas into the central regions. 
 However, more recent 
studies show that circumbinary accretion may impart additional angular momentum on the binary and eventually lead to 
the expansion of the binary separation \citep{2017MNRAS.466.1170M, 2019ApJ...871...84M, 2019ApJ...875...66M, 2019arXiv191105506D, 2020ApJ...889..114M}.  How broadly applicable these results are to astrophysical binaries is not yet clear.

Recoil velocities depend strongly on pre-merger spins and spin orientations \citep{2007PhRvL..98w1101G,2007PhRvL..98w1102C,2008PhRvD..77l4047B,2010ApJ...715.1006K,2012PhRvD..85l4049B,2012PhRvD..85h4015L,2018PhRvD..97j4049G}, 
which are poorly constrained both 
in simulations and observations.
Gas discs can crucially influence the spins. The interaction of the disc with MBH spin 
happens mainly via two mechanisms:
\begin{enumerate}
\item In what is known as the Bardeen-Petterson (BP) effect \citep{1975ApJ...195L..65B}, misalignment between the gas disc angular momentum and the MBH spin 
angle torques the two vectors towards alignment with each other.
\item The angular momentum 
of matter accreted onto the MBH changes the spin of the MBH \citep{1999MNRAS.305..654K}.
\end{enumerate}

Many studies implementing the BP effect find that in a 
gas-rich environment with a coherent gas flow, the MBH in a binary on average spins 
up and becomes aligned with the disc prior to merger \citep[e.g.][]{1996MNRAS.282..291S, 2007MNRAS.381.1617M,2009MNRAS.400..383M,2014MNRAS.441.1408T} and, as a result it experiences smaller recoil velocities 
\citep{2012PhRvD..85h4015L,2015MNRAS.451.3941G}. Simulations by \citet{2010MNRAS.402..682D} find that  
MBH spins align with the angular momentum of their orbit on time scales of 
 $<1-2$ Myr. They report typical alignments of $\sim 10^\circ$ ($\sim 30^\circ$) for cold (warm) discs. One-dimensional simulations reported the existence of critical configurations where the disc is expected to break, potentially leading to larger misalignment angles \citep{2014MNRAS.441.1408T,2020MNRAS.496.3060G}.  
However, spinning up of the MBH might not always be the case. In the case of chaotic accretion, where the matter inflow comes 
from different directions and at different speeds, the different accretion efficiencies between prograde and retrograde orbits will, on average, spin the black holes down \citep{2006MNRAS.373L..90K,2017MNRAS.465.2643C}.
In addition to that even in smooth gas flows, outer annuli can torque inner annuli leading to wild fluctuations in the spin misalignment \citep{2012MNRAS.425.1121H}. 

If a MBH binary has misaligned spin when it enters the GW dominated regime, the spin 
orientation will be modified by relativistic spin precession.
At orbital separations $a\gg GM/c^2$, where $M$ stands for the total mass of the MBH binary, the system can be
studied in the Post-Newtonian (PN) approximation \citep[e.g.,][]{2014LRR....17....2B,2016gwdw.book....9W}. MBH spins precess and orbital energy is lost to GWs on timescales proportional to $a^{5/2}$ and $a^{4}$, respectively
\citep{1994PhRvD..49.6274A}. 
At separations $a\sim GM/c^2$, the PN approximation breaks down and systems need to be studied using full numerical-relativity 
simulations \citep[e.g.][]{2014ARA&A..52..661L}. 

We utilize data from the cosmological hydrodynamic simulation suite 
Illustris \citep[e.g.]{2014MNRAS.444.1518V, 2014Natur.509..177V, 2014MNRAS.445..175G, 2015A&C....13...12N}. 
The Illustris simulation has successfully reproduced many 
of the observed properties of galaxies and their MBHs, such as galaxy merger rates, 
stellar and MBH mass functions, the cosmic star formation rate density and the baryonic 
Tully-Fisher relation \citep{2014MNRAS.444.1518V, 2014MNRAS.445..175G, 2015MNRAS.452..575S}. 
It has also been extensively used for studies of recoiling MBH and MBH binary evolution 
\citep{2016MNRAS.456..961B, 2017MNRAS.464.3131K, 2017MNRAS.471.4508K,2018MNRAS.477..964K,2020MNRAS.491.2301K}.

In this paper, we focus on modeling and characterizing the spin evolution of MBHs in a cosmological framework and studying its effects on MBH mergers and recoil velocities.
In particular, we study the dependence of spins and recoils on parameters such as the MBH accretion 
rate and the orbital eccentricity. 
We also explore how these effects may impact the number of precessing binaries 
observable by LISA. 
In this work we model gas and GW driven binary MBH spin evolution in a cosmological framework.
Our model predicts MBH merger rates with important implications 
for hierarchical structure formation and galaxy-MBH coevolution.

In Sec.~\ref{secmodel} of this paper we provide a description of the model. In Sec.~\ref{secresults} we 
discuss our findings, 
including the dependence of MBH binary spin misalignment on initial spin distributions, 
accretion rates, and eccentricities. We also examine the resulting recoil velocity 
distributions, as well as the fraction of binaries that should be strongly precessing 
in the LISA band.
In Sec.~\ref{secdisc} we discuss our conclusions.

\section{Description of the Model}
\label{secmodel}

For our analysis we use data from the Illustris 
project\footnote{\url{http://www.illustris-project.org/}}, which is a 
cosmological hydrodynamics simulation suite that reproduces key
observables of galaxy and active galactic nucleus (AGN) populations over cosmic time.
Because our focus is on MBH evolution, we primarily utilize the masses, accretion
rates, and redshifts of merging MBHs. 
To evolve binary inspiral below the simulation resolution, we follow the prescription put forward by 
\citet{2017MNRAS.464.3131K,2017MNRAS.471.4508K}, where extrapolated density profiles of the host galaxies are used to estimate the MBH hardening rates in the DF, LC, and CBD stages. 
The GW dominated regime is modeled using the PN framework implemented in the \mbox{\textsc{precession}} 
code \citep{2016PhRvD..93l4066G}. We use PN 
evolution up to separations of $a=10 GM/c^2$ where we apply fitting formulae derived
from numerical-relativity simulations to estimate the properties, including the recoil, of the merger remnant. In order to account for statistical robustness, we have run 10 different realizations of each model. Throughout this paper we denote the mass of the heavier MBH with $m_1$, the mass of the lighter companion with $m_2$, the binary total mass with $M =m_1+m_2$, and the mass ratio with $q=m_2/m_1
\leq 1$.

\begin{figure}
 \begin{center}
  \includegraphics[height=2.4in]{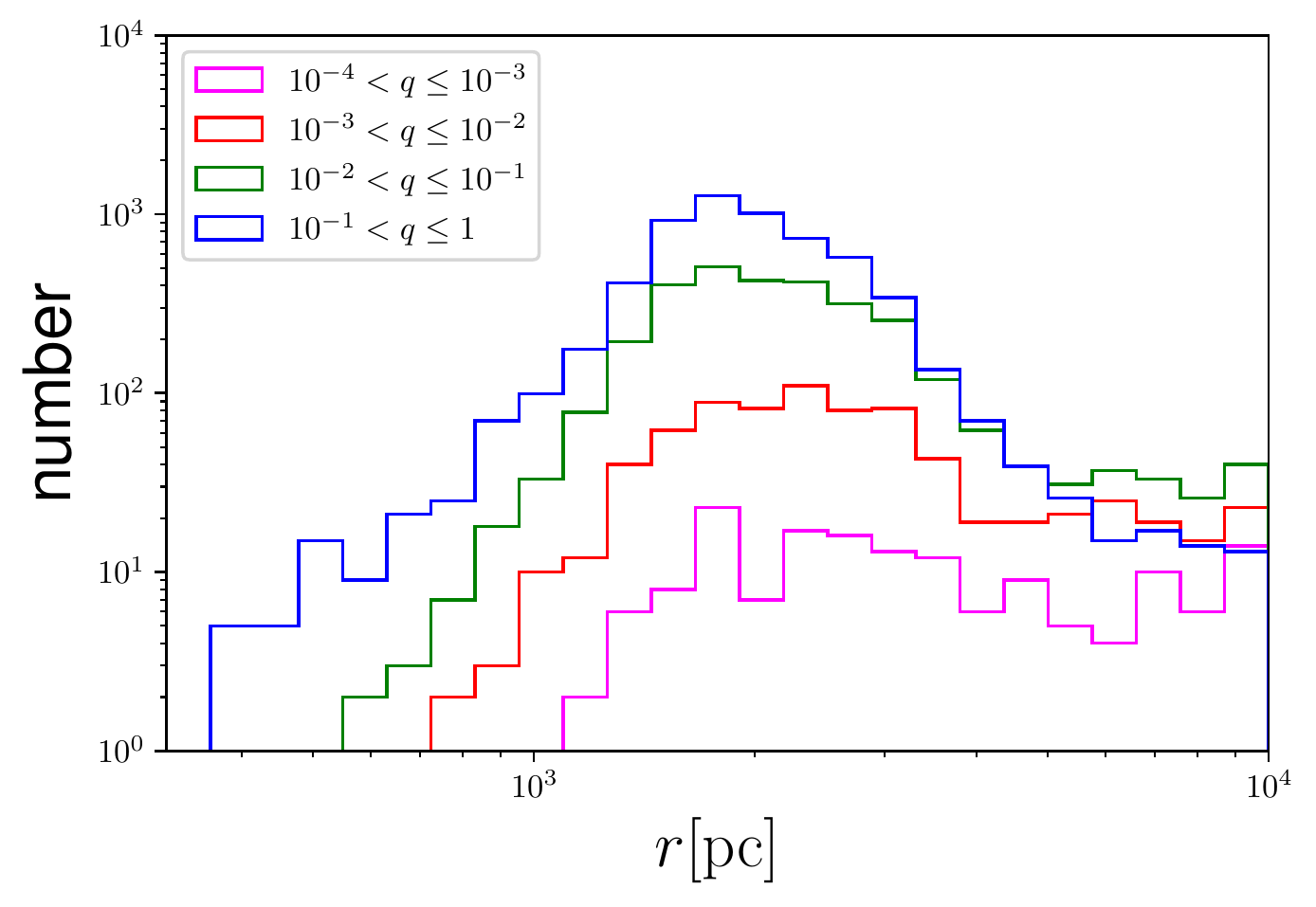}
  \caption{Distribution of gravitational softening lengths for MBH binaries in Illustris. 
  For each binary the maximum of the softening lengths is taken as the initial binary separation. In the Illustris simulations, MBHs instantaneously merge as they get within a particle softening length of each other. These softening lengths, 
even at the very small tail, represent values of the order of a few hundred pc while the GW-driven regime takes place on mpc scales.  
At these separation binaries are far from merged. For some of the binairies  it can take more than a Hubble time to go from these large separations down to GW dominated radii and coalescence depending on the host properties. Therefore a sub-resolution model is needed in order to understand the binary evolution below the softening lengths where evolution is not resolved by Illustris. 
} 
  \label{bh_hsml}%
 \end{center}
\end{figure}

\subsection{Illustris simulation suite}

Cosmological hydrodynamic simulations generally use one of two approaches: (i) smooth particle hydrodynamics (SPH)
\citep[e.g.][]{1977MNRAS.181..375G, 1977AJ.....82.1013L} 
or (ii) an Eulerian mesh-based approach \citep[e.g.][]{1989JCoPh..82...64B}. 
The Illustris simulation leverages the \textsc{arepo} code \citep{2010MNRAS.401..791S} which combines 
the advantages of both Eulerian and SPH approaches  
based on an unstructured moving mesh. The mesh is formed from a 
Voronoi tesellation based on a set of discrete mesh-generating seeds that can freely move and create
a dynamic topology  \citep{2010MNRAS.401..791S}.

Particles represent stars, dark matter (DM) and 
massive MBHs \citep{2013MNRAS.436.3031V, 2014MNRAS.444.1518V, 2014Natur.509..177V}. The MBH particles
in Illustris are seeded at a mass of \mbox{$1.42\times 10^5$ \msun} and placed in all halos
that have at least a mass of $7.1\times 10^{10}$ \msun\ and lack a MBH \citep{2015MNRAS.452..575S}. The 
algorithm assigns the highest density
gas particle as the MBH and places it at the minimum of the halo potential. 
After formation, MBHs can grow either through Eddington-limited Bondi accretion or mergers 
\citep{2005MNRAS.361..776S, 2005Natur.433..604D}.
When two MBHs come to within a gravitational softening length of each other, they are merged instantaneously. Computational requirements imply that a gravitational softening length is typically around a few kpc (see Fig. \ref{bh_hsml}) 
where, in reality, MBHs are still far from merger. 
Our main focus here is to understand and model the evolution of MBHs and their spins on these sub-resolution 
scales.

Illustris, like many comological simulations, uses a repositioning scheme to stabilize the MBH dynamics, 
wherein the MBH is always placed onto the potential minimum of its host halo. Especially for unequal-mass 
mergers, this might cause MBHs in small satellite halos to merge with the larger central MBH on 
unphysically short timescales. As this primarily affects MBHs near the seed mass, we choose to exclude
the population of MBHs with a mass of $M_{\bullet} < 10^6$ \msun\   for each indivudal MBH (\citealt{2016MNRAS.456..961B}; cf.~\citealt{2020MNRAS.491.2301K}). 

The Illustris simulations are run on a cosmological box of side $L_{box}=75h^{-1}$Mpc. Throughout this 
paper we use the highest-resolution 
run, `Illustris-1'. Simulations assumes a WMAP9 cosmology with parameters $\Omega_{m}=0.2865$, $\Omega_{\Lambda}=0.7135$, $\sigma_8=0.820$, and $H_0=70.4$ \kms\ Mpc$^{-1}$ \citep{2013ApJS..208...19H}.

\subsection{Binary inspiral time scales}

The merger of the MBHs in Illustris marks the initial point of our sub-resolution, post-processing analysis. With our post-processing we have a median inspiral time scale of $\sim 8$ Giga years for the total population. For the merged systems the median inspiral time scale is $1.6$ Gyr and for the major mergers (q>0.3) that merge by z=0 the median inspiral time scale is $1.2$ Gyr.
After Illustris merger point, we evolve the 
binaries using the prescription from \citet{2017MNRAS.464.3131K,2017MNRAS.471.4508K}. 
The binary hardening---i.e the shrinking of the binary separation---happens through four different processes:
DF, LC, interaction with CBD, and GW radiation. 
 
\begin{figure}
 \includegraphics[height=2.4in]{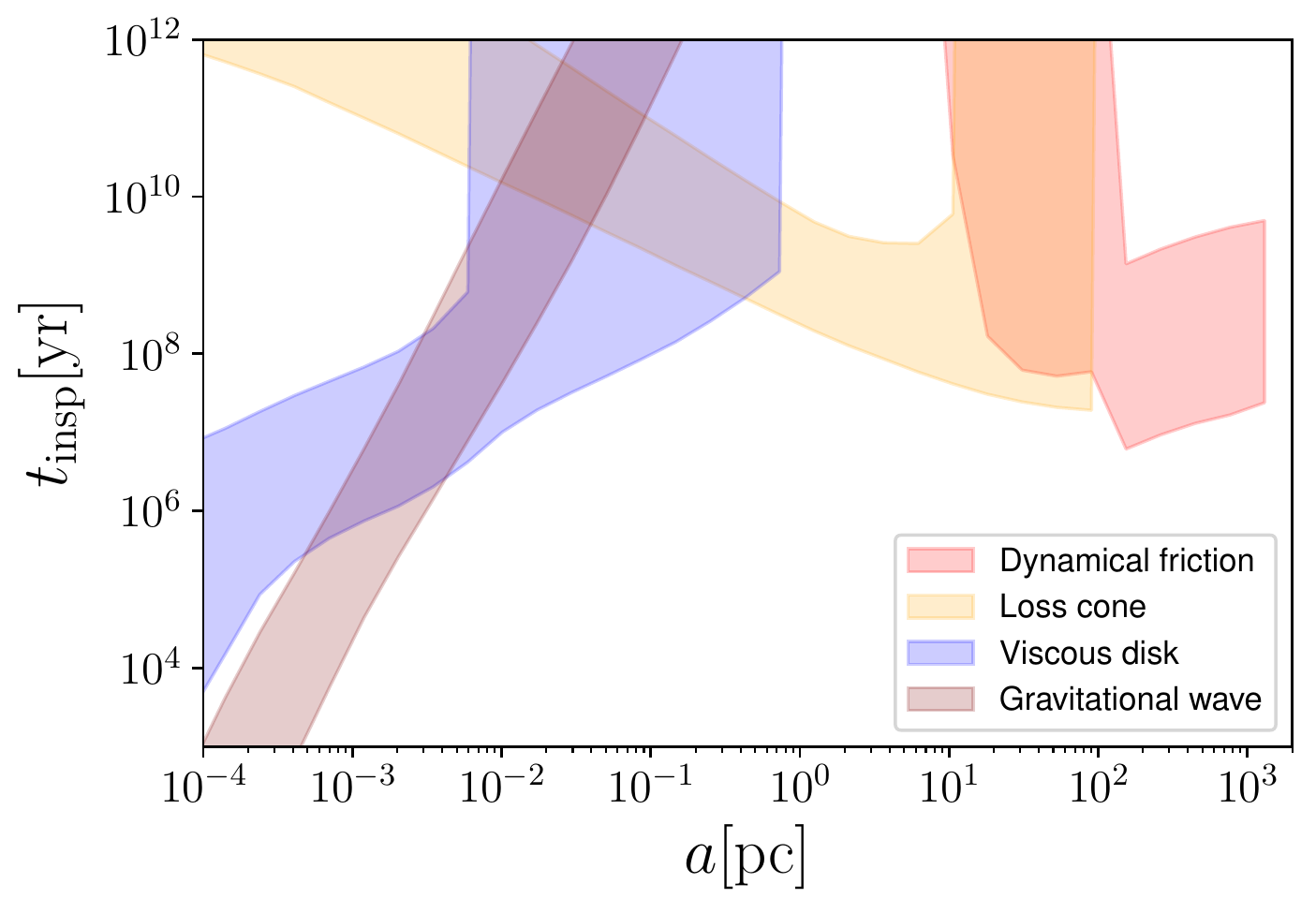}
\caption{Hardening 
time scales for different mechanisms for the 
 middle 50\% of the population. DF starts from a few kpc and dominates up to a few pc after which 
 LC takes over up to
 a few hundredth of a pc. Finally CBD and GW will dominate the inspiral at smaller separations.
 For each case the time scales are found using  $a\,({\rm d}a/{\rm d}t)^{-1}$, where $a$ is the semi-major axis of the binary.}
 \label{hardening_tscales}
\end{figure}

A moving MBH in a background of DM, gas, and stars will perturb the background by creating a
gravitational wake that removes orbital energy from MBH and thermalizes the background. 
During the early stages of galaxy coalescence, this effect, known as dynamical friction (DF), is the most dominant form of energy dissipation 
\citep{2012ApJ...745...83A, 2017MNRAS.464.3131K}. 
The DF calculation follows the change in velocity of a massive object due to an encounter with a single background particle and follows the seminal treatment by \citet{1942psd..book.....C, 1943ApJ....97..255C}.
 The hardening rate due to the DF is denoted by $(da/dt)_{\rm DF}$. The corresponding inspiral time is estimated as $(t_{\rm insp})_{\rm DF}=a(da/dt)_{\rm DF}^{-1}$. 
Figure~\ref{hardening_tscales} shows the hardening time scales due to DF in orange. In particular, we find that DF is the most dominant hardening mechanism for MBH separations larger than a few hundred pc.

From a few hundred pc to a few tenths of a pc, stellar scattering (``Loss Cone'' in Fig.~\ref{hardening_tscales}) typically dominates the MBH inspiral. At this stage, only 
low-angular-momentum stars can interact with the binary. 
Individual scattering events extract energy from the binary by ejecting the star from 
the system at high velocities.
The treatment of LC scattering in \citet{2017MNRAS.464.3131K} is based on models of tidal 
disruptions from \cite{1999MNRAS.309..447M} and scattering experiments by \cite{2008ApJ...686..432S} for circular and eccentric binaries, respectively. The LC hardening rate is denoted by $(da/dt)_{\rm LC}$. The LC 
hardening rates and hardening time scales for our population of binaries are shown in 
Fig.~\ref{hardening_tscales} in yellow.

Generally, if there is enough gas, hardening through LC continues until gas accretion onto 
the MBH binary increases significantly and a circumbinary disc forms. At this stage CBD hardening can  
become the dominant mechanism through which the binary loses 
energy 
\citep{1980Natur.287..307B, 2000ApJ...532L..29G, 2005ApJ...630..152E}. The change in binary separation in the CBD phase is denoted by $(da/dt)_{\rm CBD}$. This effect can be further enhanced following a 
galaxy merger event where a significant amount of gas is drawn into the center of the potential.
In addition to fuelling accretion onto the MBHs, the CBD phase can drive the  rapid inspiral of 
the binary up to the GW dominated regime.
Our CBD hardening rate is based on the thin-disc model of \citet{1973A&A....24..337S} and follows 
the prescription by \citet{2009ApJ...700.1952H}. In particular, we adopt the basic picture of a 
binary in a thin circumbinary disc such that the plane of the disc is aligned with the 
binary orbit. The disc gas density which enters the hardening rate is extracted directly from the accretion rate of the remnant MBH in Illustris 
\citep{2017MNRAS.464.3131K}. 

Accretion rates in Illustris are determined according to the 
Bondi-Hoyle prescription, capped at the Eddington limit.
As described in \citet{2013MNRAS.436.3031V} and \citet{2015MNRAS.452..575S}, the accretion rates calculated in the Illustris simulations are derived self-consistently with thermal, radiative, and radio-mode 
feedback models.  In designing the simulations, a small number of free parameters are calibrated to ensure agreement with observations.  In particular, by design, the population of MBH from Illustris accurately reproduce the observed masses of MBH in the local universe and also the observed luminosity function of AGN and quasars.  Taking these together means that the accretion rates in Illustris are broadly consistent with observations.

The details of the accretion process on small distance scales are unresolved in cosmological simulations.  Additionally, the timescales typically associated with `steady-state' accretion disks are also unresolved.  The behavior in Illustris can only appropriately be compared to long-duration steady-states in which the local accretion rate (and disk structure) must be consistent with large-scale gas feeding.  When accretion rates are low (e.g.~$\lambda_\textrm{Edd} \ll 1$) 
 the dynamical impact of the disk is also negligible.  Because the simulations also enforce an Eddington limit, the relevant range of accretion rates ($\lambda_\textrm{Edd} \lesssim 1$) are consistent with a thin disk. Thus, throughout our post-processing analysis of sub-resolution scales, disk surface densities are calculated based on the thin-disk assumption and the accretion rates from Illustris.

The hardening rates and hardening time scales for the CBD stage are shown in 
Fig.~\ref{hardening_tscales} in blue. For the CBD stage the outer-edge of the disk is limited by the Toomre stability criterion \citep[as calculated in][]{2009ApJ...700.1952H}.

At separations below a few hundredths of a pc, the binary loses energy mostly through emission
of GWs. The rate at which the orbital separation decreases due to gravitational radiation is given by \citep{1964PhRv..136.1224P}: 

\begin{equation}
 \left(\frac{d a}{dt}\right)_{\rm GW}=-\frac{64G^3}{5c^5} \frac{m_1 m_2 \left	(m_1 +m_2\right)}{a^3} \frac{(1+ 73 e^2/24+ 37e^4/96)}
 {(1-e^2)^{7/2}}.
 \label{dadt_gw}
\end{equation}

where $e$ is the orbital eccentricity. 
The GW hardening time scales are estimated as  $(t_{\rm insp})_{\rm GW}=a ({\rm d}a/{\rm d} t)_{\rm GW}^{-1}$,  and are shown in brown in Fig.~\ref{hardening_tscales}.

\subsection{Gas-driven spin evolution}
\label{gasdrivenspin}

A key dynamical effect of the CBD phase 
is the evolution of MBH spin angular momenta. We study only prograde orbits in the CBD phase, as the complex dynamics that may arise in retrograde CBDs are poorly understood and beyond the scope of this work.  The alignment of the individual MBHs with their corresponding discs happens 
through accretion and relativistic Lens-Thirring precession; this is referred to as the BP effect \citep{1975ApJ...195L..65B}. The MBH spins align with the angular momentum of the inner disc relatively quickly (on the viscous time) while 
the outer region remains misaligned, creating a warped profile.
The shear forces in the warped inner region will eventually align the outer and inner 
regions of the disc \citep{1996MNRAS.282..291S,2006MNRAS.368.1196L,2007MNRAS.381.1617M,2020MNRAS.496.3060G}. The time it takes
for the outer and inner discs to align with each other is given by \citep{1996MNRAS.282..291S, 1998ApJ...506L..97N, 2013MNRAS.429L..30L}:
\begin{equation}
 t_{\rm al} \simeq 3.4 \, \alpha \frac{M}{\dot{M}} \bigg(\frac{\chi}{\alpha_2}\frac{H}{R}\bigg).
 \label{alignment_time_scale}
\end{equation}

Here $M$ is the MBH mass, $\dot{M}$ 
is the accretion rate, $\chi$ is the dimensionless spin parameter, $\alpha$ is the
\cite{1973A&A....24..337S} viscosity parameter, $\alpha_2$ is the vertical viscosity coefficient, and $H/R$ is the aspect ratio of the disc. For our fiducial model we assume $\alpha=0.1$ and $H/R=10^{-3}$. The value  $\alpha_2\simeq 5.34$ is approximated using the small-warp approximation \citep{1999MNRAS.304..557O}.

Tracking in detail the variation of MBH binary spins with time is beyond the scope of this work. Rather, we identify the systems most likely to remain misaligned when they enter the GW regime by comparing the inspiral and alignment timescales in the CBD phase. 
Once the spin alignment time scale $t_{\rm al}$ is calculated, we must compare it with the 
inspiral time scales evaluated at the disc radius to determine the degree of misalignment before 
GW emission takes over. 
The effective gas disc radius $r_{\rm disk}$ 
is estimated by comparing the 
CBD hardening rate to all other rates and determining the location where CBD becomes the dominant process. In other words, $r_{\rm disk}$ is defined as the largest separation at which $\dot{a}_{\rm CBD}>\dot{a}_i$ where $i$ stands for DF, LC, and GW. This prescription gives us disk radii that range from $\sim10^{-3}-1$pc. 
 If the BP spin alignment time is longer than the gas-driven inspiral timescale, we assume a `misaligned' spin distribution at the start of the GW regime, and in the opposite case we assume an `aligned' distribution, described below.

The total number of binaries in our analysis is 9234, and this prescription 
yields 19\% (1723 binaries) binaries without a CBD-dominated phase. 
The median value of the total gas fraction of the galaxies hosting these binaries in the Illustris simulation 
is $\sim$0.33, while the gas dominated binaries have a median gas fraction of 
$\sim$0.43. Gas fraction is defined as the ratio of the gas mass over gas and stellar mass and its estimated at the time of spontaneous merger in the Illustris simulation. 
Gas dominated binaries tend to have a density profile that allows them to accrete more. The smaller accretion rate
in binaries with no CBD-dominated phase means the BP spin alignment is unlikely to work efficiently. For simplicity, we model them as having an isotropic spin distribution. 
For the rest of the population, we find the spin distribution by comparing their alignment time scales with the corresponding total inspiral time scale:

\begin{equation}
t_{\rm insp} = \frac{a}{\dot{a}_{\rm tot}}
\end{equation}
where $\dot{a}_{\rm tot} = \dot{a}_{\rm DF} + \dot{a}_{\rm LC}+ \dot{a}_{\rm CBD}+ \dot{a}_{\rm GW}$.

\begin{figure}
\centering
\includegraphics[height=2.2in]{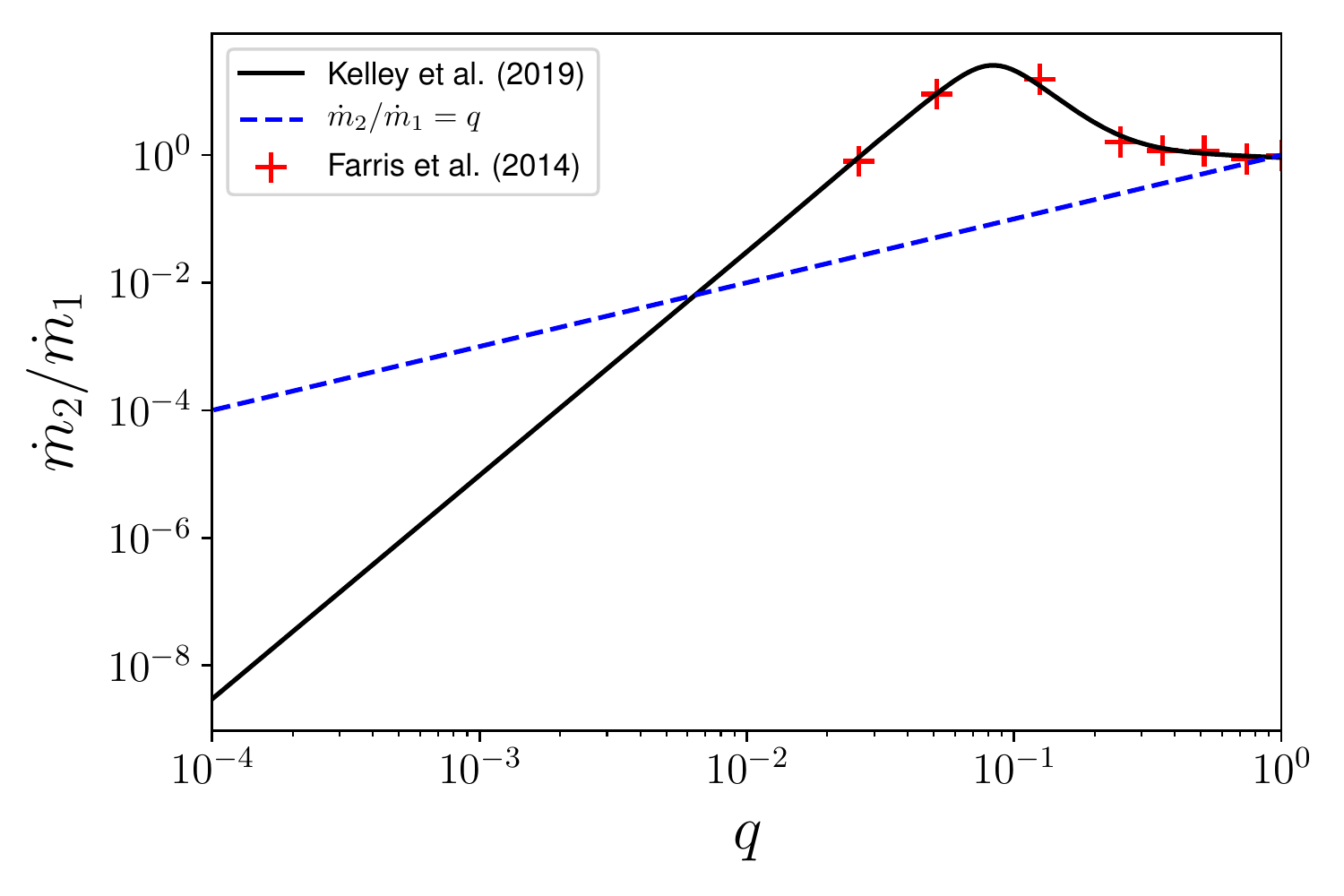}
\caption{
In a circumbinary disc, the differential accretion onto the primary and the secondary MBH is modeled using numerical simulations by   \citet{2014ApJ...783..134F} (red crosses) as fitted by \citet{2019MNRAS.485.1579K} (dashed curve, Eq.~(\ref{luke_fit})). The relative 
accretion rate between the primary and the secondary MBH has a strong dependence on the mass ratio $q$. For more extreme mass ratios, the secondary MBH orbits closer to the edge of the cavity therefore accreting most of the incoming material. Mass ratios closer to unity will reduce the asymmetry and lead to similar accretion rates onto both holes. 
Accretion rates above the blue dashed line will act to symmetrize the binary. }
\label{prim-secon-accretion}
\end{figure}

Accretion rates onto the individual MBH are not resolved by the 
Illustris simulation; 
only the accretion onto the combined binary system $\dot{M}_{\rm bin}$ are available. 
Upon formation of a circumbinary disc, the torques 
from the binary can create a gap in the circumbinary disc with a mass pile up on the inner edges of the disc. 
The mass that is accreted onto the gap will then accrete onto the MBHs, creating 
circumprimary and circumsecondary discs.

\begin{figure*}
\begin{center}
    
\begin{minipage}[b]{.4\textwidth}
\begin{center}
    
\includegraphics[height=2.4in]{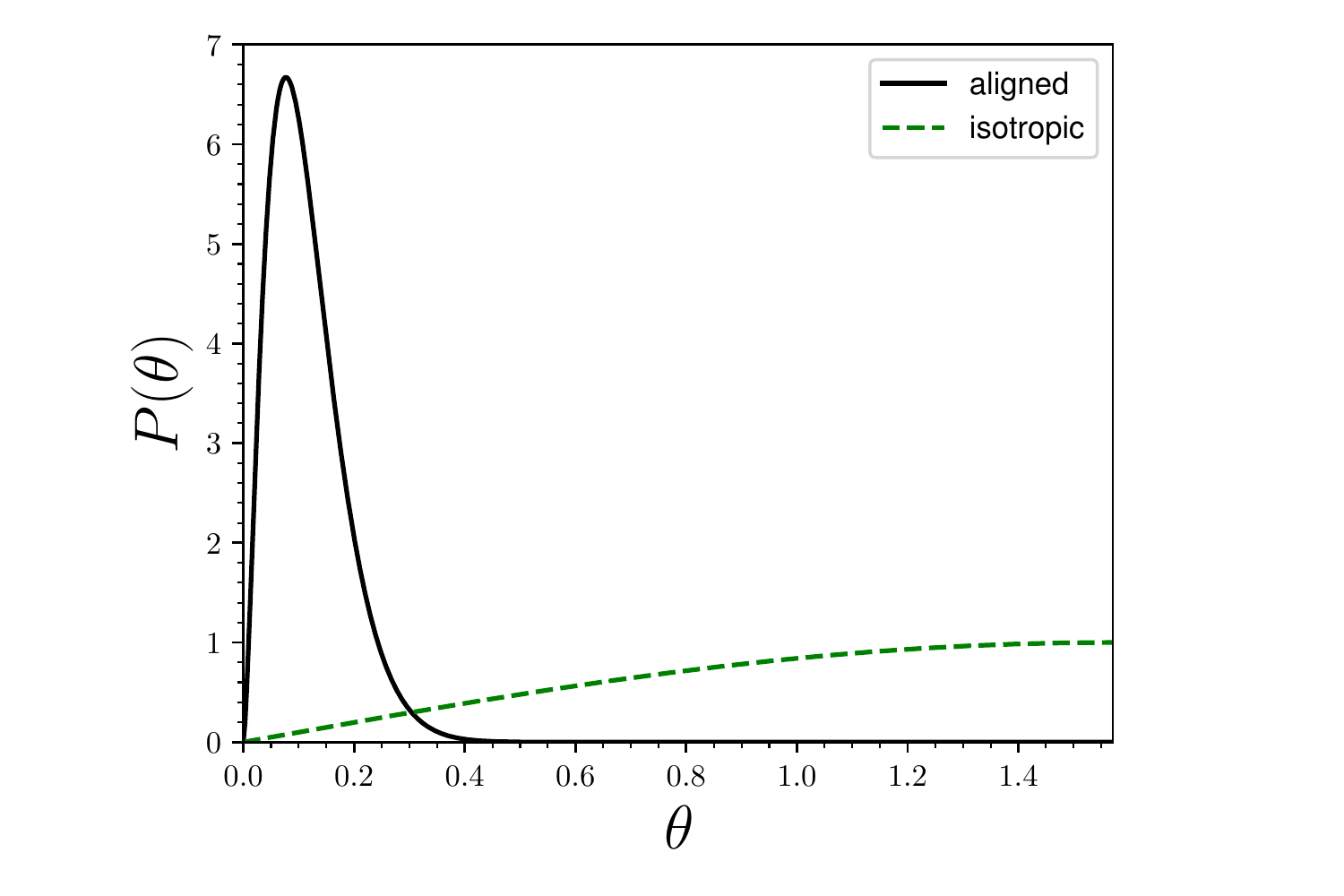}
\end{center}
\end{minipage}\qquad
\begin{minipage}[b]{.4\textwidth}
\includegraphics[height=2.4in]{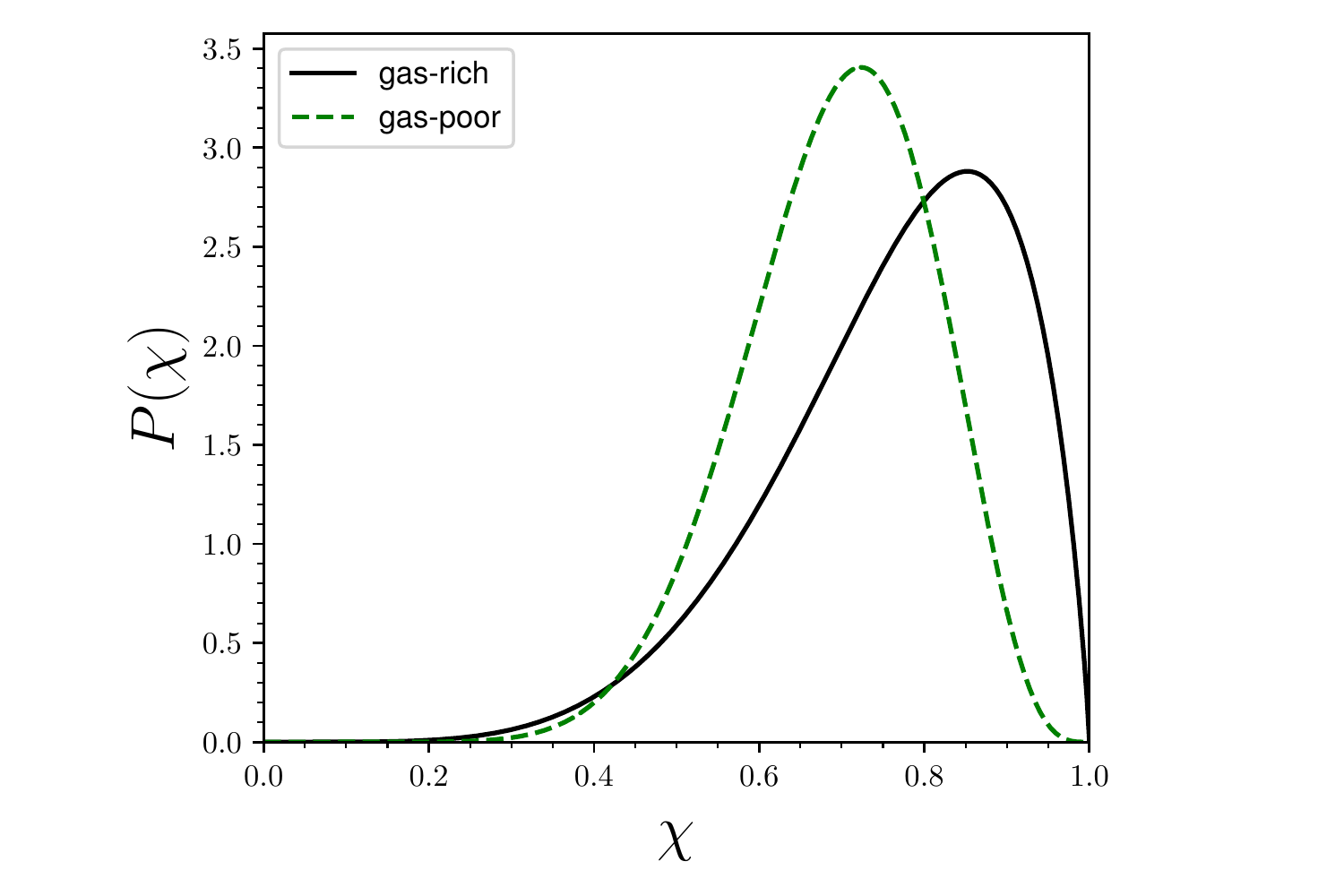}

\end{minipage}
\end{center}
\caption{Spin orientation (left) and magnitude (right) distributions used in this paper. 
For the aligned distribution \citep[Cf.][]{2010MNRAS.402..682D},  accretion is assumed to be efficient and spins are close to aligned with the orbital angular momentum.
 The isotropic distribution, representing successive dry mergers, assigns random spin orientations \citep{2016MNRAS.456..961B}.  
 The fiducial model uses a 
 combination of these, to assign spin directions, based on a comparison between inspiral and spin-alignment time scales
  Coherent accretion is also assumed to spin up the MBHs to relatively high spin magnitudes depending on the gas richness of the host. In the fiducial model dimensionless spin parameters is extracted from beta distributions that peak around 0.7 and 0.8 for gas-poor and gas-rich halos. Gas richness of the halo is based on the gas fraction which is defined as the ratio of the gas mass over total baryonic mass.  
 If the alignment time scale is larger than the inspiral time scales, we assume that the MBH remains misaligned 
 by the end of inspiral, and in the fiducial model these spin directions  are assigned from the 
 isotropic distribution. When inspiral time scales are larger than 
 alignment time scales, the MBHs are assigned spin directions 
 from the aligned distribution. The assignment of the spin magnitudes $\chi$ (i.e. dimensionless spin parameter) is based on the gas fraction of the remnant halo. For gas fractions above and below 0.2 gas-poor and gas-rich distributions are assigned respectively. }
\label{fig:all_dist}
 \end{figure*}

Simulations of the dynamics of gas accretion onto the binary have shown a strong correlation between the accretion
rates and mass ratio $q$	 \citep{1994MNRAS.268...13S,2008ApJ...672...83M,2008ApJ...682.1134H,
2013MNRAS.436.2997D,2014ApJ...783..134F,2017MNRAS.466.1170M,2019ApJ...871...84M}. For 
small mass ratios, the less massive secondary will orbit closer to the edge of cavity and 
clear away most of the matter falling
into the gap. However if the mass ratio is too small ($q\lesssim 0.03$) the secondary's 
accretion will not be 
strong enough to curb the primary's accretion rate \citep[cf.][]{2019arXiv191105506D}. Therefore accretion onto the primary is 
favored for $q\to 0$. Our models neglect possible modulations in accretion rate due to non-zero eccentricity, as discussed by e.g. \citet{2016ApJ...827...43M}.
For larger mass ratios, symmetry implies that matter falls 
roughly equally onto each MBH.
These combined effects 
have been referred to as ``differential accretion'' \citep{2015MNRAS.451.3941G}.

The ratio $\dot{m}_2/\dot{m}_1$ of the accretion rates is estimated using simulations by 
\citet{2014ApJ...783..134F} (red crosses in Fig. \ref{prim-secon-accretion}) as fitted by \citet{2019MNRAS.485.1579K} (dashed line):
\begin{equation}
  \frac{\dot{m}_2}{\dot{m}_1}={q^{a_1}e^{-a_2/q}+\frac{a_3}{(a_4q)^{a_5}\;+\;(a_4q)^{-a_5}}},
 \label{luke_fit}
 \end{equation}
where $a_1=-0.25$, $a_2=0.1$, $a_3=50$, $a_4=12$, and $a_5=3.5$ (cf.~\citealt{2015MNRAS.451.3941G} for a different fit).
We assume that all of the matter from the cirumbinary disc accretes onto either of the two MBH, i.e. $\dot{M}_{\rm bin}=\dot{m}_1+\dot{m}_2$ (but see \citealt{2013MNRAS.436.2997D,2016MNRAS.460.1243R}). 
The individual accretion rates for primary and secondary found here are used in Eq.~\ref{alignment_time_scale}
to evaluate the alignment time scales. 

Following the CBD evolution, the MBHs reach the final stage of merger, which
is dominated by GW emission.
The spin distributions found by differential accretion constitute the initial conditions for our PN integrations. 	

\subsection{Spin distributions}
\label{spindistsec}

The spins of the MBHs prior to merger strongly depend on the accretion rates in the CBD phase \citep[see][ for more on the accretion and spin relations]{2005ApJ...620...59S, 2005ApJ...620...69V, 2012MNRAS.423.2533B} . 
A binary with a high accretion rate in a gas driven phase
will affect the alignment of the spins with the disc through the Bardeen-Petterson effect, 
leading to a higher degree of alignment of spins with the orbital angular momentum vector.
In particular, the spin magnitude will increase as  
${ d\mathbf{\chi}/{dt}}\propto \dot{M}$ .
It is important to note that the time scales at which the spin magnitude changes are much larger than the time scales
for spin alignment \citep{2005ApJ...620...69V}. This is because in the BP effect, spin alignment is set by the dynamics of the disc warped region, while the spin magnitudes rate is set by the material flowing through the BH innermost stable circular orbit. 
 Also, for a significant change in the spin magnitude, 
the MBH needs to accrete of the order of its own mass \citep{1970Natur.226...64B,1999MNRAS.305..654K}.  Given these considerations we do not evolve the spin magnitude of the MBHs in this work.

Let us denote with $\theta_1$ and $\theta_2$ the angles 
between the MBH spins and the orbital angular momentum of the disk. Note that we assume both of the MBH are lying on the plane of the disk. 
The different timescales involved imply that we cannot use the same prescription for spin alignment and spin magnitudes.

The distribution of $\chi$ is informed by the host galaxy properties. Specifically, we use the total gas fraction of the remnant 
galaxy assuming that a higher gas fraction will lead to a more coherent flow that spins up the MBH. the gas fraction is 
defined as the ratio of the gas mass to the total baryonic mass of the galaxy. 
We develop two distributions for $\chi$, which we dub as  ``gas-rich'' and ``gas-poor'' as shown in the right panel of Fig.~\ref{fig:all_dist}. 
The ``gas-poor''  case represents a scenario that could be due to the MBH going through successive dry 
mergers with randomly-oriented spins. In this case, the dimensionless spin parameter is extracted from a 
beta distribution that peaks at $\sim$0.7 \citep{2016MNRAS.456..961B}.
On the other hand, the ``gas-rich``scenario represents a case 
where accretion is more efficient at spinning up the MBH. In the ``gas-rich'' case the dimensionless spin parameter is extracted from a beta distribution that peaks at $\sim 0.8$. 

We choose a critical gas 
fraction of $0.2$ as our gas richness criterion. MBH binaries in halos with higher gas fractions are assigned spin magnitudes based on the ``gas-rich''  distribution, and MBH spin magnitudes in gas-poor halos are 
assigned based on the ``gas-poor'' distribution.
 While this choice is arbitrary, we find that our  results do not depend strongly on this choice. In other words a population that is fully assigned a ``gas-rich'' distribution or a ``gas-poor'' distribution to its spin magnitudes give very similar misalignment percentages and recoil velocity curves.

We also develop two distributions ``aligned'' and ``isotropic''for the spin directions $\theta_i$. These distributions are shown on the left panel of Fig.~\ref{fig:all_dist}. 
In the ``aligned'' case accretion is more coherent and the spins are more closely aligned with the orbital angular momentum vector \citep{2010MNRAS.402..682D}. On the other hand the ``isotropic'' case represent dry mergers with less efficient spin alignment. In our analysis we are not evolving the spin vectors but rather using a time scale analysis to assign distributions.
For the spin directions we compare inspiral and alignment timescale at $r_{\rm disk}$ and assign spin direction based on them. The following is a summary of our model:

\begin{eqnarray}
\label{ModelII}
 \left\{ 
    \begin{array}{rl}
      P(\chi): & %
    \begin{array}{rllll} 
      & f_{\rm gas}>0.2 & & {\rm gas\mbox{-}rich} \\
      & f_{\rm gas}<0.2 & & {\rm gas\mbox{-}poor} 
    \end{array}
      \\
      \\
      P(\theta): & %
    \begin{array}{rllll} 
      & t_{\rm al} > t_{\rm insp} & & {\rm misaligned} \\
      & t_{\rm al} < t_{\rm insp} & & {\rm aligned} 
    \end{array}
      \\
    \end{array}
    \right.
\end{eqnarray}
  \newline
$P(\chi)$ and $P(\theta)$ denote the $\chi$ and $\theta$ distributions. $f_{\rm gas}$ indicates the gas fraction of the host halo. $t_{\rm insp}$ and $t_{\rm al}$ are inspiral and alignment time scales, respectively, in the gas-driven inspiral phase. Our distributions for both spin magnitude and directions are shown in Fig.~\ref{fig:all_dist}.

\begin{figure}

 \includegraphics[height=2.3in]{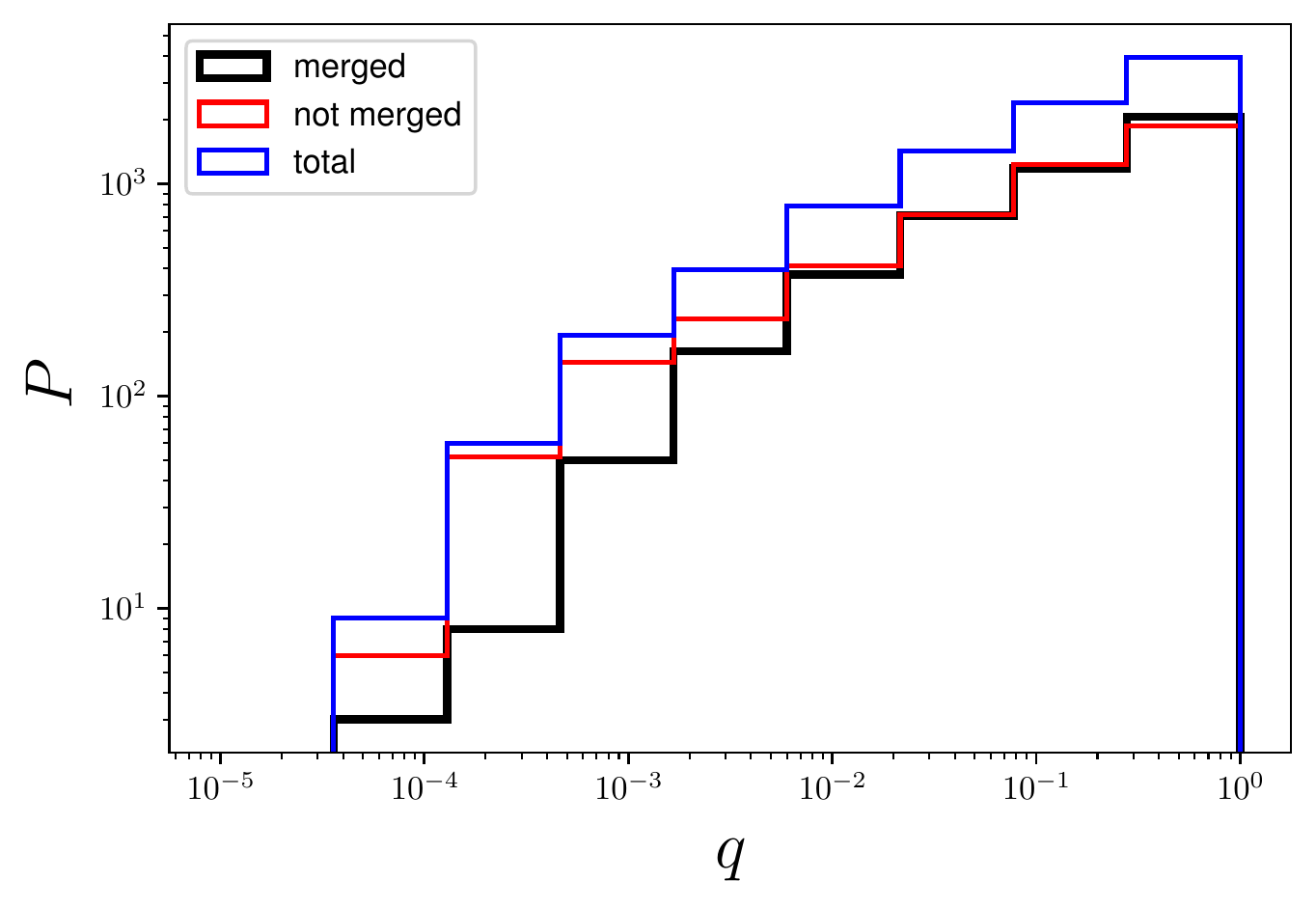}
 \caption{Distribution of mass ratios $q$ for the total population (blue), as well as the merged (black) and non-merged (red) binaries by $z=0$. The merged 
population is made of 4451 binaries out of a total sample of 9234. The merged population is
skewed towards higher mass ratios.}
 \label{merged_not_merged_q}
\end{figure}

\subsection{Gravitational-wave driven evolution}

In the GW-dominated stage, we follow the binary evolution using a post-Newtonian (PN) approach. 
We make use of the \textsc{python} module \textsc{precession} \citep{2016PhRvD..93l4066G}. 
In particular, precession-averaged integrations \citep{2015PhRvL.114h1103K,2015PhRvD..92f4016G} allows us to evolve the binary orbital angular momentum and the BH spins directly from the large separations predicted by the previous CBD or LC phase. The code assumes black-hole binaries on circular orbits. The treatment is accurate up to 2PN in spin precession and 1.5PN in radiation reaction. Integrations are initialized at the separations where GWs start dominating the hardening rate. 
Precession-averaged evolution is performed down to a final separation of $a=10 GM/c^2$ at which the final angles $\theta_i$ and $\Delta\Phi$ are calculated assuming random precessional phases.
(the spin magnitudes are constant to very high PN order; see \citealt{2015PhRvL.114h1103K,2015PhRvD..92f4016G} for details). The 
initial
values of $\theta_1$ and $\theta_2$ are provided by the previous (typically CBD) phase, while the 
initial angle $\Delta\Phi$ between the spin components in the orbital plane
is assumed to be randomly distributed in $[-\pi,\pi]$. 

Following the precession averaged evolution the final values of the parameters are used 
to evaluate fitting formulae to 
numerical relativity simulation and estimate the properties  of the post-merger MBH. In particular, the recoil fit is calibrated on simulations by \cite{2007ApJ...659L...5C,2007PhRvL..98i1101G,2008PhRvD..77d4028L,2013PhRvD..87h4027L,2012PhRvD..85h4015L} as collected by \cite{2016PhRvD..93l4066G}.

\section{Results}
\label{secresults}
\subsection{Fiducial model}

\begin{figure} 
\includegraphics[height=2.4in]{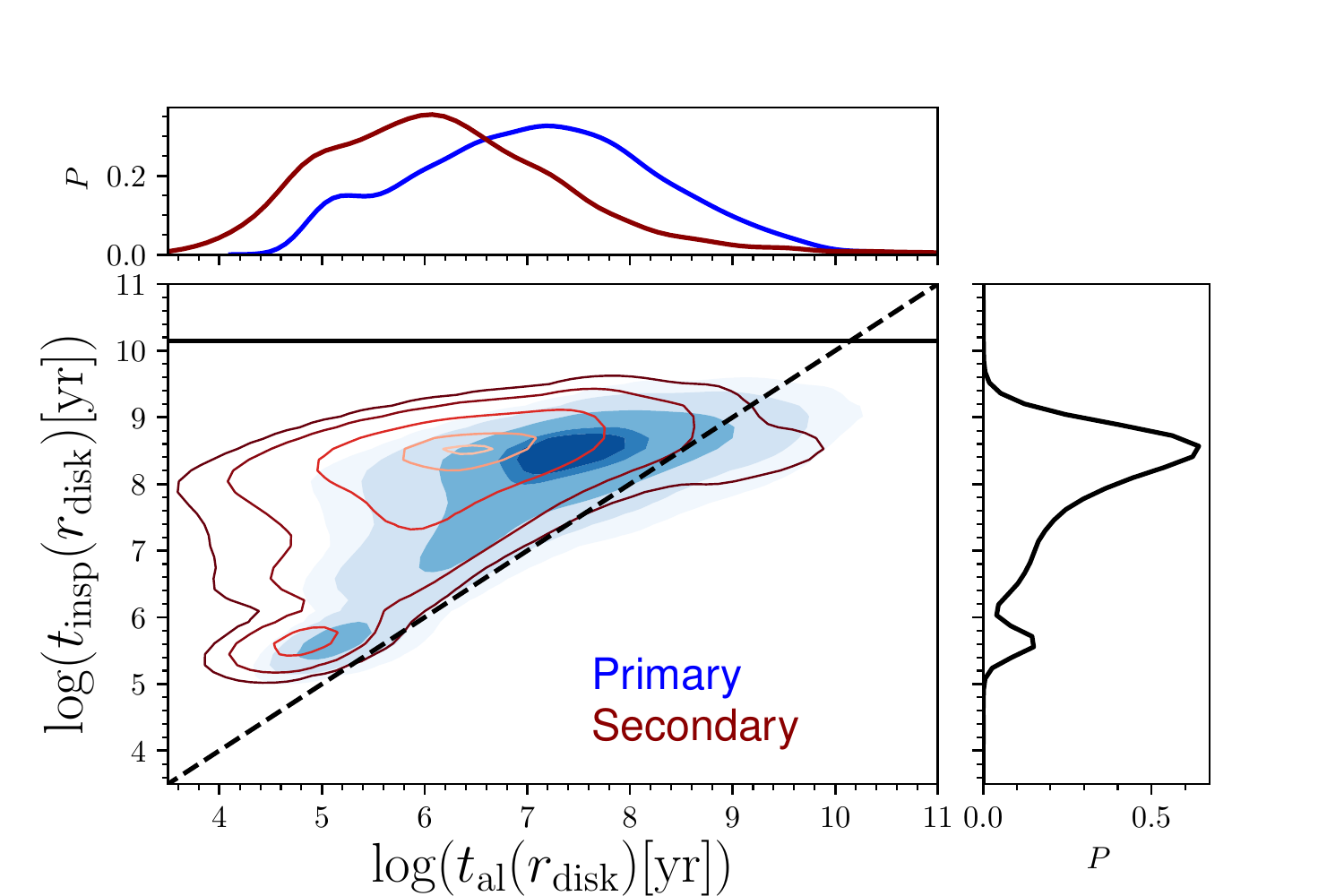}
 \caption{Gas-driven inspiral and BP alignment time scales for the fiducial model, calculated from the point at which gas-driven inspiral begins to dominate the binary hardening ($r_{\rm disk}$). The blue filled and red line contours 
indicate primary and secondary MBH's alignment time scales and they correspond to 99\% (outermost contour), 97\%, 90\% , 80\% , 77\% (innermost contour)  of the probability distribution function,  respectively. Differential accretion along with the smaller mass of the secondaries, imply that primaries take longer than the secondaries to align. 
The solid horizontal line indicates the Hubble time, while the dashed line denotes equal alignment and inspiral 
times. In our fiducial model, most MBH spins are aligned by the end of the gas-driven phase, but a non-negligible fraction remain misaligned as they enter the GW-dominated phase. 
}
\label{insp-al-tscl}
\end{figure}

Figure \ref{insp-al-tscl} shows the gas-driven MBH binary inspiral versus spin alignment time scales, 
calculated at the start of the CBD phase ($r_{\rm disk}$). 
We assume all binaries 
have an initial eccentricity of 0.6 in the fiducial model; our treatment of fixed initial eccentricities follows that in \citet{2017MNRAS.471.4508K,2018MNRAS.477..964K} . The eccentricity is assigned at beginning of DF phase, however, it only changes in LC and GW dominated phases in our model. Eccentricity would also greatly affect accretion onto the MBH binary and the differential accretion but we don't take this into account in our model. 
Nevertheless, the choice of eccentricity does not significantly affect our final result, as discussed in Sec.~\ref{eccsec}. In the GW-dominated phase, the hardening rate is strongly dependent on the eccentricity: $t_{\rm insp} \propto (1-e^2)^{7/2}$, see Eq.~\ref{dadt_gw}.
Higher eccentricities
 will in principle enhance the GW hardening rates 
 and reduce the time to MBH merger.  
However, in our fiducial model with initial eccentricity of 
$0.6$, we find only $1.7$\% binaries, that do not have a GW dominated phase.
These rare binaries all accrete at the Eddington rate in their final stages and have unusually high CBD hardening rates; some also have unusually low GW hardening rates. 

Figure \ref{merged_not_merged_q} shows the mass ratio for merging and non-merging MBH binaries in our model. During the evolution we calculate the redshift at each step of evolution and the merged binaries are the ones that merge by redshift z=0. The ones that don't merge have inspiral time scales larger than a Hubble time. The binaries that don't merge are omitted in the GW regime since they don't contribute to the merger rate. Thus they are not contributing to LISA merger rates either.
Figure \ref{merged_not_merged_q} also shows that the mass retios for the merged population is skewed towards larger mass ratios ($q>0.1$). This combined with the differential accretion (Sec.~\ref{gasdrivenspin}), implies that the accretion rate is 
typically dominated by the secondary MBH.  This leads to larger misalignment time scales for the primary, as seen in Fig.~\ref{insp-al-tscl}. Given the smaller mass of the secondary, with the higher accretion rates caused by differential accretion, its  spin alignment is further enhanced. In particular, we find that 
19\% of the primaries and 10\% of the secondaries are misaligned at the end of the CBD phase. 
 Differential accretion in the CBD phase can also drive the binary towards $q=1$. However, the total accretion in the CBD phase is not enough to significantly change the mass ratio distribution \citep[cf.][]{2020MNRAS.498..537S}.
Therefore, we make the simplifying assumption of constant mass ratios.

\begin{figure}
\hspace{-3em}
\includegraphics[height=2.7in]{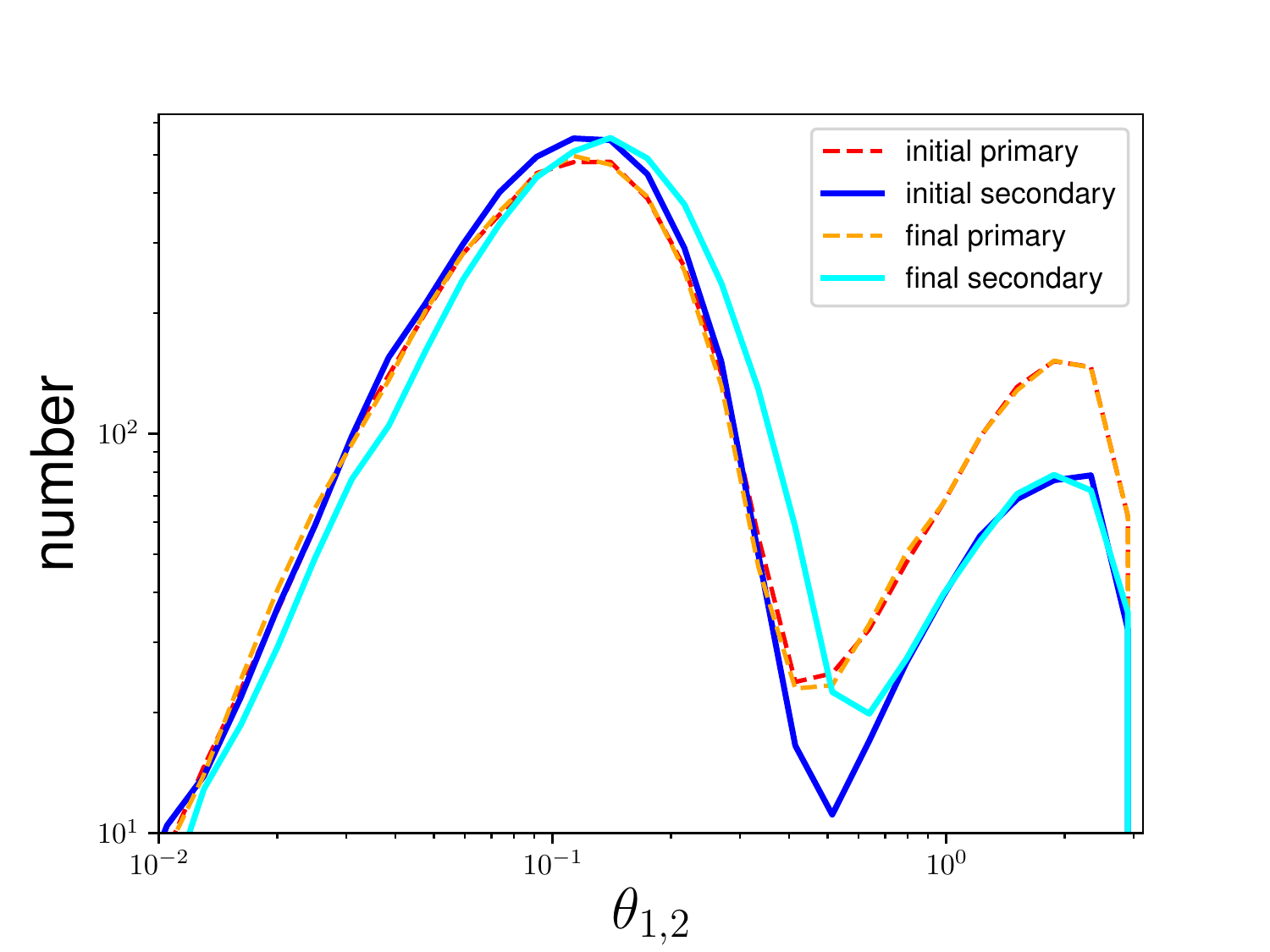}

 \caption{Initial and final angles 
 between MBH spins and the binary orbital angular momentum in the GW-dominated phase, for both primary (red, yellow) and secondary (blue, cyan) MBHs.
 These polar angles $\theta_{1,2}$ are initialized at 
 the start of the GW phase based on a comparison between inspiral and alignment time scales, as shown in Fig. \ref{insp-al-tscl}. For misaligned spins 
 ($t_{\rm al}>t_{\rm insp}$), spin orientations are assigned from an isotropic distribution, and for $\rm t_{al}<t_{insp}$, 
 spins are assigned from the "aligned" (partially-aligned) distribution as discussed in 
 Sec.~\ref{spindistsec}. 
 Although GR precession can induce large changes in individual spin angles, the overall distribution remains similar, with a slight increase in the misalignment of the initially aligned secondary MBH spins.  
 }
 \label{thetai_prim_sec}
\end{figure}

The comparison between inspiral and alignment timescales (Fig.~\ref{insp-al-tscl}) dictates the configuration of spin 
orientations at the onset of the GW phase. 
This effect can be seen in the ``initial'' configurations in Fig.~\ref{thetai_prim_sec}, which shows that primary MBHs (red curve) are, on average, more misaligned than secondaries (blue curve).

These distributions  of $\theta_{1}$ and $\theta_{2}$, 
along with an isotropic distribution of the angle $\Delta\Phi$ between the spin components in the orbital plane,
provide the initial conditions to track the BH spins in the GW-driven phase. As shown in  Fig.~\ref{thetai_prim_sec}, the distribution of polar angles for the primary MBHs does not change significantly during this phase. Its important to point out that individual spins can and do 
change greatly,
even though the overall distribution varies little.
Relativistic spin-spin couplings imply that the behavior of the secondary MBH spin is affected by the primary MBH spin. In systems 
where the primary MBH spin is misaligned with the orbital angular momentum, relativistic precession tends to induce 
greater misalignment in the secondary. Conversely, if the primary MBH spin is nearly aligned and the secondary 
is misaligned, spin precession tends to drive the secondary into greater alignment. 
In other words, the trend is such that as the separation of angles decreases: the configuration of polar angles tend to go
towards $\cos\rm{\theta_1}\simeq \cos\rm\theta_2$ (cf. \citealt{2004PhRvD..70l4020S,2013PhRvD..87j4028G,2020PhRvD.101l4037M}).
Isotropic spin distributions tend to 
remain isotropic during 
this phase \citep{2007ApJ...661L.147B}. The anisotropic 
distributions, however, are 
more significantly affected by relativistic precession \citep{2004PhRvD..70l4020S,2010PhRvD..81h4054K,
2015PhRvD..92f4016G,2015PhRvL.114h1103K} where the modification of angles before the merger is stronger. 

\begin{table*}
\begin{center}
\begin{tabular}{ lllrrrrrrr } 
\hline
  \textbf{host } & \textbf{$\mathbf{\theta_1(r_{\rm GW})}$} & \textbf{$\mathbf{\theta_2(r_{\rm GW})}$} & \textbf{median $q$} & \textbf{median $M_{\rm bin}$} & \textbf{\% of mergers} & \textbf{median $\mathbf{\rm v_{\rm p}}$} &\textbf{\% v>500\kms} & \textbf{\% v>1000\kms} \\ 
  & & & &$[M_{\odot}]$ & \textbf{(merged binaries)} &\textbf{[\kms]} &\textbf{[\kms]} & & \\
  \hline

  {\bf Merged binaries} & {\bf fiducial}    & {\bf fiducial}  &  {\bf 0.22} & {\bf 4.7 $\mathbf{\times 10^7}$}& {\bf 100} & {\bf 147} & {\bf 12} & {\bf 2.6}   \\
  Gas rich  & isotropic  & isotropic & 0.59  & 6.4 $\times 10^6$& 0.54 & 711 &    65${\pm9.7}$  & 34${\pm8.8}$    \\ 
  Gas rich  & isotropic  & aligned   & 0.12  & 1.8 $\times 10^7$& 2.4  & 248 &  28${\pm2.9}$  & 12${\pm2.6}$   \\ 
  Gas rich  & aligned    & isotropic & --    & -- & 0.0  & --  &  --  & --     \\ 
  Gas rich  & aligned    & aligned   & 0.36  & 1.9 $\times 10^7$ & 27   & 189 &  14${\pm1.2}$  & 1.8${\pm0.4}$   \\ 
  Gas poor  & isotropic  & isotropic & 0.042 & 1.1 $\times 10^9$ & 8.1  & 42.8 & 21${\pm1.1}$  & 9.8${\pm1.2}$   \\ 
  Gas poor  & isotropic  & aligned   & 0.077 & 1.1 $\times 10^8$ & 7.9  & 111  &  10${\pm1.1}$  & 3.1${\pm1.3}$   \\ 
  Gas poor  & aligned    & isotropic & 0.002 & 1.5 $\times 10^9$ & 1.4  & 0.07 & 0.0${\pm0}$ & 0.0${\pm0}$   \\ 
  Gas poor  & aligned    & aligned   & 0.24  &  5.4$\times 10^7$& 52   & 136  & 10${\pm0.7}$  & 1.2${\pm0.2}$  \\ 

  \hline
\end{tabular}
\end{center}
\caption{
GW recoil velocity statistics are listed here for our fiducial model (first row in bold) and the eight sub-populations that it is consisted of. Data in this table show percentages evaluated from merged binaries only. The first column denotes whether the host galaxy is gas rich or gas poor (as defined above); this designation determines the initial assignment of BH spin magnitudes in our calculation. The second \& third columns distinguish systems in which each BH is aligned or not aligned by the end of the gas-driven phase ($t_{\rm align} < t_{\rm insp}$ vs. $t_{\rm align} > t_{\rm insp}$); this determines whether each BH is assigned a spin orientation from the ``aligned'' or ``isotropic'' distribution. The fourth \& fifth columns indicate the median mass ratio $q$ and median binary mass $M_{\rm bin}$ for each sub-population. The sixth column indicates the percentage of merging binaries that fall into each sub-population. The seventh column indicate their median recoil velocity with precession ($v_{\rm np}$). The eighth \& ninth columns give the percentage of each sub-population resulting in recoil kicks above 500 and 1000 \kms, respectively. Note that binaries in gas-rich hosts have more equal mass ratios than those in gas-poor hosts, resulting in somewhat higher recoil velocities for the former. We can also see here that misaligned primaries contribute more to higher recoil velocities. 
Values are averaged over 10 realizations, with the standard deviations shown for the kick-velocity percentiles in the final two columns.} 
\label{fiducial spin models}
\end{table*}

\begin{figure}
    \centering
    \includegraphics[height=2.4in]{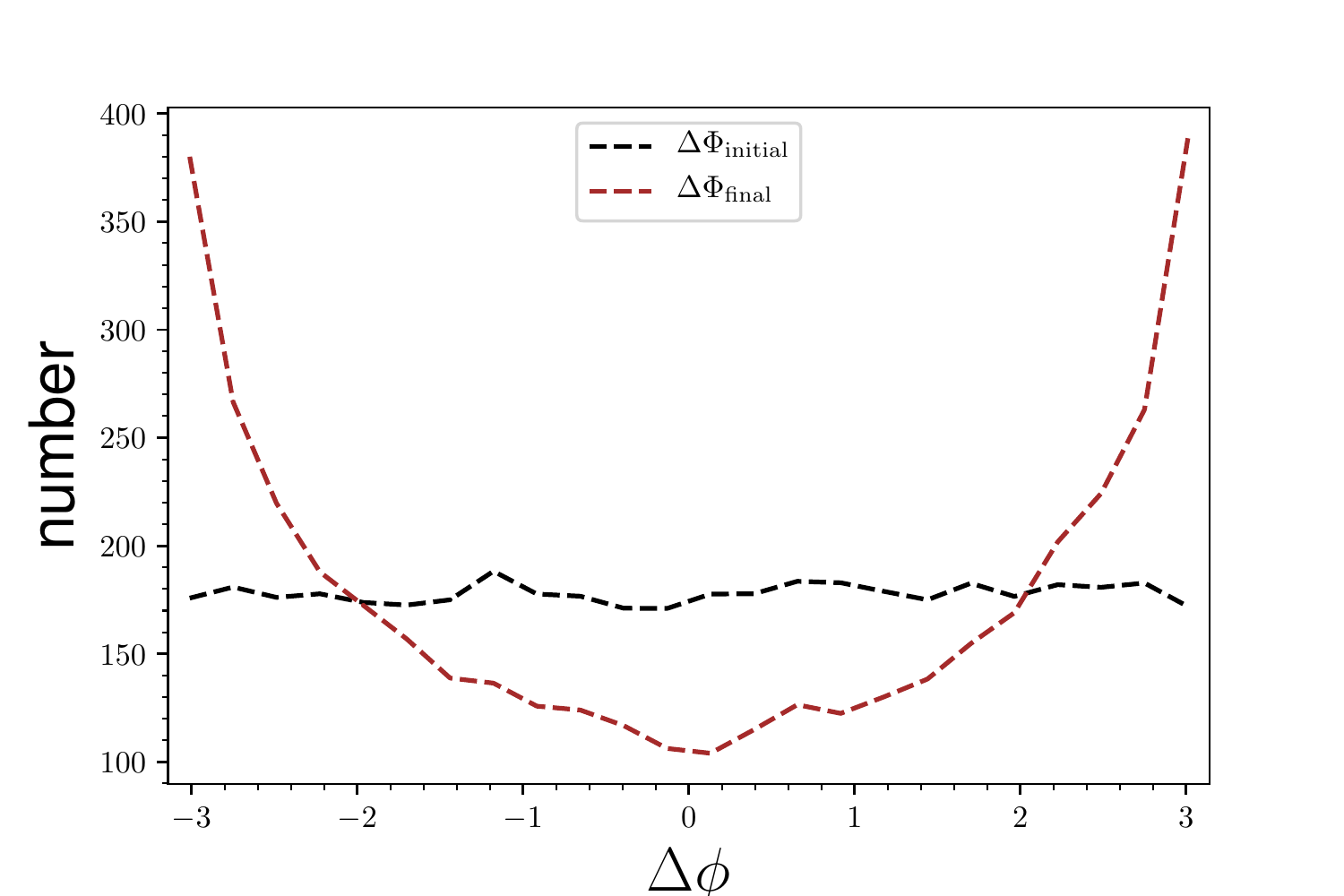}
    \caption{
    Initial and final distributions of the angle $\Delta\Phi$ between the spin components in the orbital plane during the GW-dominated phase. The initial distribution is determined at the onset of GW-dominated phase (i.e. end of disk phase). Since the disk phase does not affect $\Delta\Phi$, its distribution at the beginning of GW phase is isotropic.  However, in the GW-dominated phase, MBH spin precession drives the distribution towards $\Delta\Phi=\pm \pi$ when the secondary spin is aligned, as is the case for 90\% of the mergers listed in Table~\ref{fiducial spin models}.
    }
    \label{deltaphi}
\end{figure}

Although spin precession does not dramatically impact the distribution of polar angles $\theta_1$ and $\theta_2$ for our fiducial model, it does strongly affect the distribution of differences in azimuthal angles $\Delta\Phi$ as shown in Figure \ref{deltaphi} \citep[see][]{2010ApJ...715.1006K}.  This occurs because the BP effect aligns the secondary spin in 90.02\% of our mergers, and MBHs with aligned spins and mass ratios $q \lesssim 0.5$ are preferentially driven into the $\Delta\Phi = \pm\pi$ librating spin morphology during the GW-driven phase of the inspiral \citep{2015PhRvD..92f4016G}.

MBHs in this librating spin morphology should have higher kicks because they are closer to the asymmetric "superkick" configuration \citep{2007PhRvL..98w1102C}, but we find that precession has a negligible effect on the median recoils for the eight sub-populations in Table~\ref{fiducial spin models}.  This may be an artifact of the bimodal distributions of the spin directions $\theta_i$ in Fig.~\ref{thetai_prim_sec}; $\Delta\Phi$ is undefined and thus irrelevant in the limit that one or both of the spins is aligned, while distributions in which both spins are initially isotropic remain isotropic throughout the GW-driven phase \citep{2007ApJ...661L.147B}.

However, precession can significantly affect individual velocities \citep{2010ApJ...715.1006K,2020arXiv200501747R}. The precession-induced changes in recoils $|V_{\rm p} -V_{\rm np}|$ (where 
`p' stands for precession and `np' stands for no precession), can reach $\sim 3000$ km/s for individual cases. This is consistent with the known sinusoidal variation found in numerical-relativity simulations of ``superkicks'' \citep{2008PhRvD..77l4047B,2018PhRvD..97j4049G}. Around 52\% of the merging population experiences an increase in velocity due to precession, and the rest experience a decrease in recoil velocity due to precession. More specifically, 71\% of our MBH present recoils  that change by at least 
10 \kms\ when precession is accounted for 
, 34\% of recoil velocities change by at least 100 \kms, and only 0.7\% change by more than 1000 \kms. 
 Table \ref{fiducial spin models} shows recoil velocity distributions for the different sub-population in our model. As expected the highest recoil velocities happen for the gas rich and isotropic spins. The velocities in the gas rich model are higher because this subset of binaries is consisted of systems with higher mass ratios compared to the gas poor subset. In the gas poor subset we have higher median MBH masses. This means a robust LC hardening that makes the binary merge before a Hubble time. 

The recoil velocity distribution for our fiducial model is shown in Fig.~\ref{fig:recoil_accr_all}. For comparison, we also show the velocity distribution that results from assigning spins to all MBHs from the ``aligned'' distribution, and from the ``isotropic" distribution (cf. Sec.~\ref{spindistsec}). For all of the distributions in the figure the spin magnitude, $\chi$, assignment  follows the gas dependent criterion given in Eq. \ref{ModelII}. If 
we assume the ``aligned" distribution, spins are nearly aligned with each other and the orbital angular momentum at
the onset of the GW phase. In this case, the recoil velocity distribution peaks at $\sim$ 140 \kms. On the other hand, 
for the ``isotropic'' distribution, the recoil velocity 
distribution peaks at $\sim$700 \kms, with a large tail of recoils $>$1000 \kms.
Based on our fiducial model, the misaligned portion
of the population, 19\% of primaries and 10\% of secondaries, are assigned a random spin orientation 
and the rest are assigned spins from the ``aligned'' distribution. Therefore, the recoil 
velocities in the fiducial model 
lie between a purely ``aligned'' and purely ``isotropic'' distribution, as shown in Fig.~\ref{fig:recoil_accr_all}. 
While the fiducial model has a recoil velocity distribution that peaks at around the same value as the aligned distribution,
it presents a higher-velocity tail that extends to $\sim$ 3000 \kms. 
There are $\sim12$\% fiducial systems with a recoil velocity of \mbox{$\geq500$ \kms} and $\sim3$\% systems with a recoil velocity of \mbox{$\geq 500$ \kms}.

\subsection{Dependence of spin evolution on accretion environment}

Because the accretion-disc scale is far smaller than the resolution of Illustris, 
the simulated Bondi accretion rates are inherently uncertain and  may well be over-estimated. 
Although accretion rates could in principle be under-estimated, they are Eddington limited and 
their distribution in Illustris is strongly peaked at the Eddington 
limit during MBH mergers \citep{2016MNRAS.456..961B}, which is where we focus on in this paper. This makes the  over-estimate scenario more likely. Combining that with the fact that our results are quite sensitive to accretion rates, we consider alternate models with lower $\dot{m}$ to determine the impact on our results, if in fact these high accretion rates are over-estimated during merger events.
To this end we have repeated our calculations with artificially reduced accretion rates by a factor of 100.
Furthermore, because a significant number of the merging MBHs in Illustris are Eddington-limited at the time of merger ({9\%}) 
, this reduced accretion model variation is effectively testing a scenario where all of these MBHs are low-luminosity rather than high-luminosity AGN. With this reduction factor, 31\% of merging MBHs have Eddington ratios  $\gtrsim 10^{-3}$, as opposed to {82\%} with the fiducial model's accretion rates, which are extracted directly from Illustris.

\begin{table*}
\begin{center}
\begin{tabular}{ rrrrrrr } 
\hline
 \textbf{Disc $H/R$}&  \textbf{Disc $\dot{M}$}   &  \multicolumn{2}{c}{{\bf\% Misaligned}} & \textbf{median $\mathbf{\rm v}$ [\kms]}& \textbf{\% v>500\,\kms}& \textbf{\% v>1000\,\kms}\\ 
   &     &  \textbf{Primary} & \textbf{Secondary}  & & & \\ 
  \hline
  {\bf 0.001}& {\bf $\dot{M_{\rm fid}}$}      & {\bf 19}  & {\bf 10} &{\bf 147} &{\bf 12.47} &{\bf 2.6} \\ 
  0.001& 0.01$\dot{M_{\rm fid}}$  & 48  & 25 & 189 & 19.68 & 6.32 \\ 
  0.01 & $\dot{M_{\rm fid}}$      & 48  & 18 & 180 & 20.40 & 7.43 \\ 
  0.01 & 0.01$\dot{M_{\rm fid}}$  & 79  & 42 & 261 & 31.28 & 14.03\\ 
  \hline
\end{tabular}
\end{center}
\caption{Fraction of MBHs with misaligned spins at the start of the GW-dominated phase for our fiducial model (first row,
in boldface) and three model variations in which we modify the accretion rate and the aspect ratio of the disc. The `primary' and `secondary' columns show the misalignment percentages of the respective MBHs. For both primary and secondary MBHs, spin evolution is strongly affected by changes in the accretion rate and disc aspect ratio. The change in accretion rates modifies both the alignment time scales and the inspiral time scales, while the change in aspect ratio modifies the alignment time scales only. 
Our fiducial model uses conservative assumptions for the accretion disc, while in other models a large majority of the MBHs are misaligned at the onset of the GW-driven phase.}
\label{accretion_aspect_ratio_table}
\end{table*}

\begin{figure*}
 \includegraphics[height=3.9in]{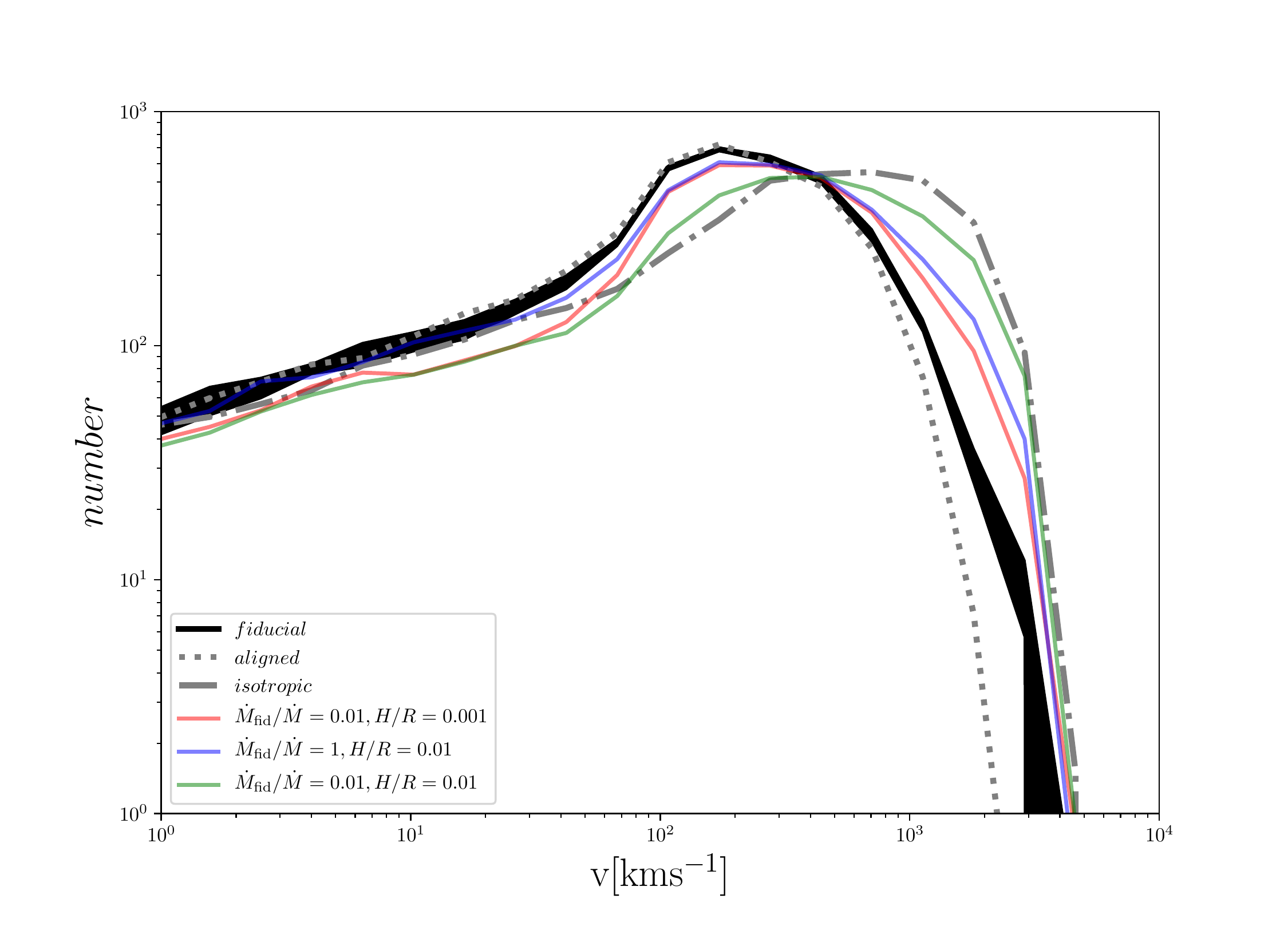}
 \caption{The black solid line indicates GW recoil velocities for the fiducial model, which assigns `aligned' or `isotropic' spin-angle distributions based on the alignment timescales. The shaded area around the fiducial model (apparent at large velocities) indicates the standard deviation over 10 realizations of our model.
 For comparison, we also include the recoil velocities resulting using entirely the `aligned' and `isotropic' distributions (grey dotted  and grey dashed dotted, respectively) along with the reduced accretion rate and thick disk models.
Greater spin misalignment of the isotropic model 
 yields higher recoil velocities while, on the other hand, for the nearly aligned distribution the recoil velocity distribution peaks at smaller values. The fiducial model, being a combination of the two distributions, sits in between 
 the two extremes, and has a tail of high-velocity recoils extending to $\sim$ 3000 \kms. With the reduced accretion model the spin orientation becomes more isotropic compared to the fiducial model. Hence the recoils are pushed to higher values. Higher disk aspect ratio is more efficient and making the distribution isotropic. Therefore, it has slightly higher kick velocities compared to reduced accretion model. Changing both the aspect ratio and the accretion rate will result in even higher kick velocities, as shown in green here.}
 \label{fig:recoil_accr_all}
\end{figure*}

Our results, shown in Table \ref{accretion_aspect_ratio_table}, demonstrate that accretion rates strongly influence
the alignment and inspiral time scales of binaries.
BP alignment time is inversely proportional to the accretion 
rate, 
and thus MBH spins will take longer to align with the disc in systems with low values of ${\dot M}$. 
In the models with reduced accretion rates, a higher fraction of binaries are misaligned when they enter the GW-driven phase ---79\% of primaries and 42\% of secondaries for the thicker disk. 
These fractions are more than three times higher than those in our fiducial model. 
As the fraction of misaligned MBHs increases, the total spin distribution will begin to resemble a isotropic distribution. 
Fig.~\ref{fig:recoil_accr_all} shows the recoil velocity for the reduced accretion model in solid brown 
and, as expected, this model shows larger recoil velocities compared to the fiducial model. We find that 19.7\% and 6.3\% of recoils are above \mbox{500} \kms\ and \mbox{1000} \kms, respectively.

Table \ref{accretion_aspect_ratio_table} also shows the dependence of alignment on the aspect ratio of the disk. 
Because the relationship between aspect ratio and accretion rate is somewhat uncertain and may depend on multiple factors \citep{1988ApJ...332..646A,1995PASP..107.1207N, 2003A&A...409..697M, 2003MNRAS.338..189M}, we vary these model components independently to span a range of possibilities.
The aspect ratio equation only enters the expression for the  alignment time scale. A smaller aspect ratio reduces the alignment time scales and hence the percentage of misaligned binaries. Table \ref{accretion_aspect_ratio_table} shows that increasing the aspect ratio from $0.001$ to $0.01$ boosts spin misalignment by more than a factor of 2 for primaries and slightly less than that for secondaries. Such a high percentage of misalignment will make the recoil distribution resemble the full ``isotropic'' case. For this model variation, we find that 20\% and 7\% of recoils are above \mbox{500} \kms\ and \mbox{1000} \kms, respectively. The recoil distribution in the large aspect ratio model has the same peak as the reduced accretion model. Finally a reduction in the accretion rates accompanied by an increase in the aspect ratio will change the distribution most significantly, by driving it closer to the ``isotropic'' distribution. With the 79\% and 42\% misaligned primaries and secondaries, respectively, the percentage of recoils above \mbox{500} \kms\ and \mbox{1000} \kms\ are 31\% and 14\%. The peak of the distribution also shifts to \mbox{$\sim500$} \kms\ compared to \mbox{$\sim150$} \kms\ for $\dot{M}_{\rm fid}/100$ and for the increased aspect ratio $H/R=0.01$.

For our fiducial model, we have also looked at the correlation of the recoil velocities with galaxy properties such as gas fraction, star formation rates, and the masses of different galaxy components (gas, dark matter, stars, and black holes).
We find that binaries that merge by $z=0$ have higher host gas fractions. Aside from this, however, the recoil velocities do not show any strong trends with the host galaxy properties. This reflects the fact that only the spin magnitudes in our model have an explicit dependence on host galaxy properties, and the difference between the ``dry-merger" and ``coherent accretion" spin magnitude distributions is relatively minor (see Fig \ref{fig:all_dist}). However, there is an important indirect connection with the host galaxies, namely that gas-poor systems have smaller mass ratios on average, as seen in Table \ref{fiducial spin models}. This suggests that in many of these cases there is a satellite merging with a more massive MBH that resides in a gas-poor elliptical galaxy. We plan to further explore the possible dependence of recoil velocities on host galaxy properties in future work. 

\subsection{Eccentricities}
\label{eccsec}

\begin{figure}
 \includegraphics[height=2.3in]{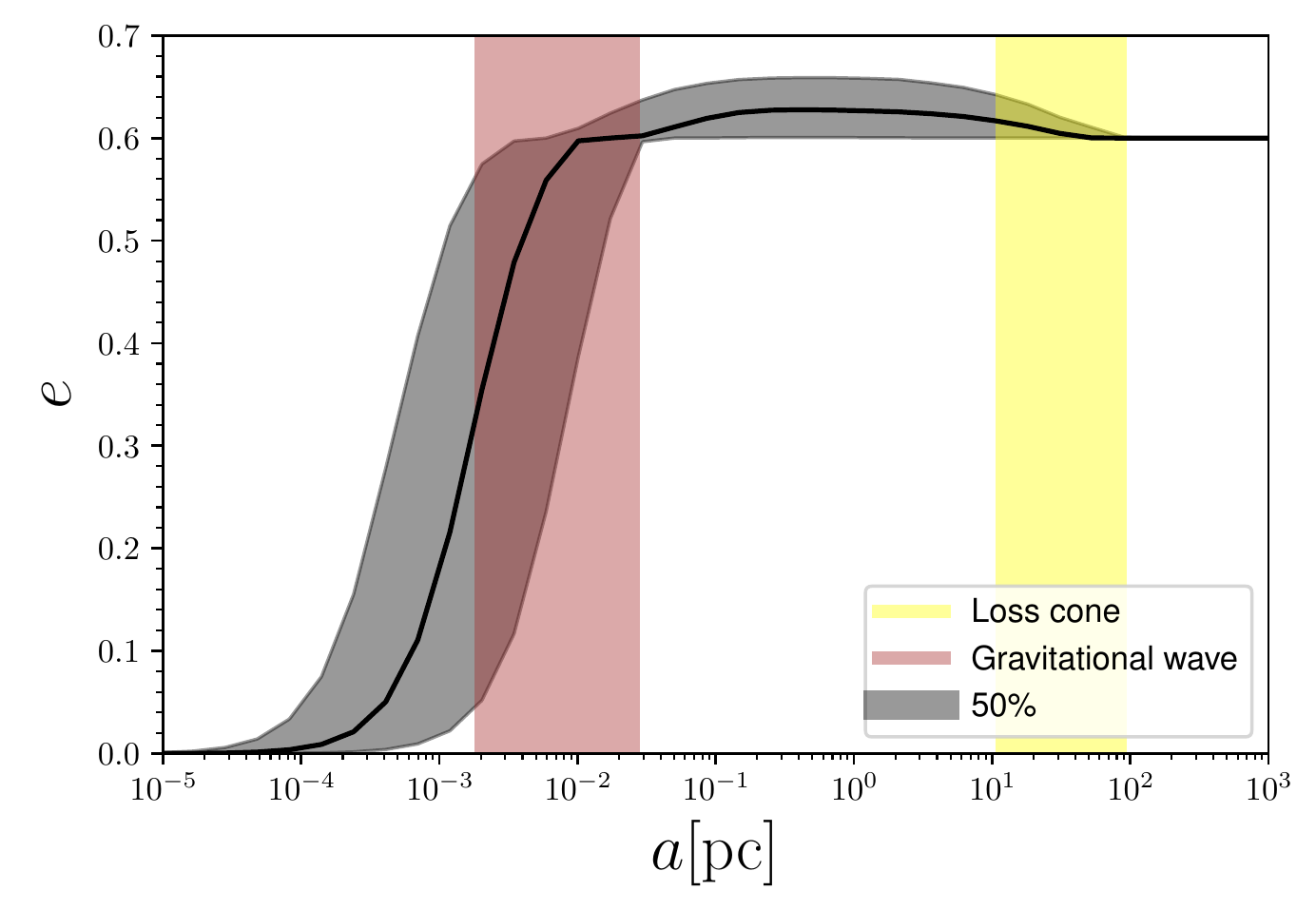}
 \caption{Evolution of binary MBH eccentricity as a function of separation. The eccentricity is evolved in the LC- and GW-dominated phases of our inspiral model. The yellow highlighted region shows the interquartile range for the radii at which binary evolution transitions from DF-dominated to LC-dominated.
 LC stellar scattering increases the eccentricity. The red highlighted region indicates the interquartile range 
 of radii at which GW hardening begins to dominate. The GW phase leads to a reduction in eccentricity and circularization of the binary.}
 \label{fig:eccentricity_evolution}
\end{figure}

In our binary-inspiral model, we can initialize the sub-resolution MBH binary orbits with non-zero 
eccentricities. 
Eccentricity is then modulated in both the LC and GW inspiral phases \citep{2017MNRAS.471.4508K} %
We do not attempt to model eccentricity evolution in the DF or CBD stages. Note also that eccentricity evolution 
is not included in the GW spin precession calculation.   
 Non-zero eccentricity at the start of the GW phase means that we should start the \texttt{PRECESSION} code at a smaller radius. In a recent study by \citep{2019PhRvD.100l4008P} this has been shown to not have a significant effect on the overall GR precession.
We can nonetheless consider the effects of precession and eccentricity evolution separately, to characterize their impact on our results.
Figure \ref{fig:eccentricity_evolution} shows how 
MBH binary eccentricity evolves during 
the LC and GW stages of evolution. 
In general, the LC phase increases the eccentricity of the binaries that are 
initially slightly eccentric or have unequal masses, while the GW phase rapidly reduces the eccentricity and circularizes the orbit
\citep{1964PhRv..136.1224P,2010ApJ...719..851S, 2013CQGra..30x4005M, 2017MNRAS.471.4508K}.

One outcome of the higher eccentricities in the LC phase
is that LC-driven inspiral 
will dominate down to smaller binary separations.
This effect marginally reduces the effective disc radii $r_{\rm disk}$ and increases the number of systems that merge without a CBD phase.
In particular, increasing the initial eccentricity at the beginning of the sub-resolution 
inspiral (beginning at the DF phase) from 0 to 0.9 increases the percentage of the systems with no CBD-dominated phase from 16\% to 25\%.

We find that varying binary eccentricities does 
not affect the distribution of recoil velocities 
in any meaningful way, primarily because 
no eccentricity evolution occurs during 
the CBD phase in our model. 
Thus, BP alignment time scales and inspiral time scales 
do not change, 
except insofar as the disc radius is modulated by eccentricity evolution in the LC-driven phase. 
The recoil velocities do not change significantly either; there is negligible change with eccentricity for $e \lesssim 0.5$, 
while for higher eccentricities, a slight increase is seen in the highest-velocity tail of the distribution.
The highest 1\% of recoil velocities are $\gtrsim 1400$ \kms\ for $e=0.5$, versus $\gtrsim 1700$ \kms\ for $e=0.8$. This comes from the more isotropic spins for the higher eccentricity. There are 18 \% and 9\% misaligned primaries and secondaries at e=0.5. for e=0.8 the misalingment percentages are 20\% and 12\% for primaries and secondaries respectively.

It is worth stressing that residual eccentricity at merger can actually be very important for black-hole recoils~\citep{2007ApJ...656L...9S,2020PhRvD.101b4044S}. Here we are neglecting those effects by construction because the numerical-relativity fitting formula we use is only valid for circular orbits. This is a good approximation because the eccentricity decays quickly before merger (Fig.~\ref{fig:eccentricity_evolution}). Eccentricity is also neglected in the spin-precession evolution. We cannot rule out the possibility that the coupled effects of eccentricity and precession could alter the final spin distribution and thus the recoil; further exploration of this is a subject for future work (see~\citealt{2019PhRvD.100l4008P}).

\subsection {MBH Merger rates}

The total merger rate, with no delay (i.e. the Illustris merger rate), for all the 9234 binaries from the simulation is \mbox{0.53 yr$^{-1}$}.
 Out of this population, 47\% (4269) merge by $z=0$ in our fiducial model with a merger rate of \mbox{0.15 yr$^{-1}$}. 
Let us recall that these rates are for MBHs with $M>10^6$ \msun\ and that  the  mass cut is implemented to avoid dynamical uncertainties regarding MBHs near the seed mass, as described in Sec.~\ref{secmodel}.
We find that the total merger rate 
does not depend significantly on the assumed initial eccentricities $e$, at the beginning of DF phase. 
The merger rates for $e=0$ and $e=0.9$ are \mbox{0.14 yr$^{-1}$} and \mbox{0.16 yr$^{-1}$}, respectively. The dependence on the accretion rate is also minimal. The reduced accretion rate model yields {0.13 yr$^{-1}$} compared to {0.15 yr$^{-1}$} for the fiducial model. 

LISA is most sensitive to 
mergers between binaries with masses $\lesssim 10^7 M_\odot$ 
 out to a redshift of $z\sim20$, with limited sensitivity to more nearby mergers at higher masses \citep[$\lesssim 10^8 M_\odot$; e.g.,][]{2016PhRvD..93b4003K,2017arXiv170200786A}. 
 We find that 67\% of the merged population (2970 binaries) 
falls within this mass range ($\lesssim 10^8 M_\odot$), 
with a corresponding  
merger rate of 
 {0.1 yr$^{-1}$}. 
The merger rates quoted 
here are not equivalent to LISA event rates, as that requires setting a detectability threshold and a consideration of the LISA noise versus binary frequency.  

Crucially,  these merger rates extracted from the Illustris MBH population will necessarily underestimate the true merger rate, primarily because our analysis is restricted to MBH masses {$\geq10^6$ \msun} 
owing to resolution limits. 
In contrast, semi-analytic models of MBH evolution, which are computationally cheaper compared to large cosmological simulations, often include prescriptions for low-mass MBH seeds \citep[$\sim 10^2 - 10^3$ \msun; e.g.,][]{2016PhRvD..93b4003K, 2016PhRvL.117j1102B}. Such models are therefore able to predict merger rates over essentially the full range of LISA sensitivity, finding merger rates as high as 23 yr$^{-1}$ \citep{2019MNRAS.486.4044B}. \citet{2019MNRAS.486.4044B} also include a model for triple MBH encounters, 
which are neglected in our analysis, 
and find that they contribute substantially to the merger rate. Note also that the efficiency of semi-analytic calculations comes at the expense of information about the internal structure of galaxies; these detailed data provided by the Illustris simulation are critical for our models of MBH binary inspiral and spin evolution.

Using Illustris binaries, \citet{2020MNRAS.491.2301K} reported a merger rate of \mbox{0.5--1 yr$^{-1}$}. They made use of a new method for dealing with the uncertainties due to the seeding mechanisms at masses $\lesssim 10^6$ \msun. \citet{2020MNRAS.491.2301K} included some, but not all, of the binaries in the mass range $10^5$--$10^6$ \msun, which we neglected 
In order to deal with the artificial mergers that were created by the Illustris Friends-of-Friends algorithm near the seed masses, \citet{2020MNRAS.491.2301K} required all merger constituents to exist for at least one snapshot before merger. They then focus on galaxies that have had their central MBH removed by the 
 re-positioning algorithm. They track the evolution of the galaxy that have lost an MBH in a flyby encounter to ensure it is not artificially seeded again. If the galaxy is seeded at some point after the encounter, that seed and all its associated mergers are removed.
MBH binaries within this mass range almost doubled their analyzed population to 17535 compared to 9234 in our analysis. 
Their results are consistent with our findings for $>10^6$ \msun.

 \citet{2016MNRAS.463..870S} presented a MBH merger analysis using the EAGLE, a large cosmological simulation with resolution and volume similar to those of Illustris \citep{2016MNRAS.457..844F}. 
  Their findings for seed masses similar to Illustris ($M_{\rm seed}=10^5$ \msun) yield about 2 mergers per year. 
Given all of the differences in the numerical techniques and sub-grid models, these results are in reasonable agreement with the Illustris merger rates. 
 
 \subsection{Characteristics of precessing binaries}

The subset of MBH binaries that undergo strong precession is of particular interest for LISA, because these systems will have the largest precession-induced modulation of their GW waveforms, which could potentially be detectable. Because the signal-to-noise ratio required to detect precession depends non-trivially on both the sensitivity curve and the amplitude of precession and nutation, we cannot directly comment on the observability of precessing binaries with LISA. Although a detailed study of precessing GW waveforms is beyond the scope of this work, here we briefly characterize the evolution of key quantities in the GR precession phase. 

\begin{figure*}
 \includegraphics[height=8.25in]{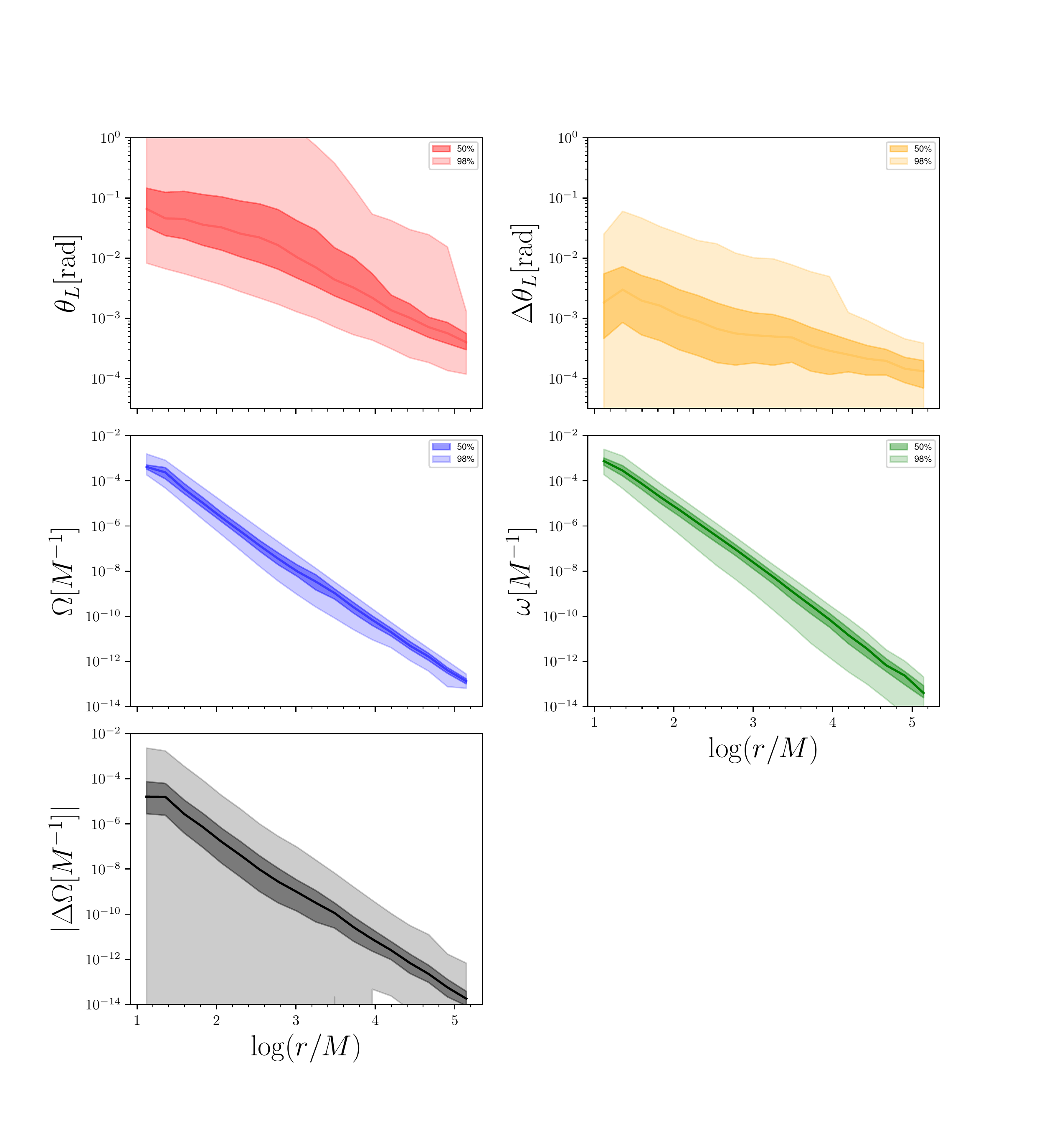}
 \caption{For our population of merging MBH binaries, the evolution of five parameters characterizing GR precession is shown.
 The five panels show the precession amplitude  $\theta_L$ (upper left), nutation amplitude $\Delta\theta_L$(upper right), precession frequency $\Omega$ (middle left), nutation frequency $\omega$ (middle right), and magnitude of oscillation of the precession frequency $|\Delta\Omega|$ (bottom panel). Angles are in radians, and frequencies are in dimensionless $M^{-1}$ units. These quantities are the key spin observables affecting modulation and amplitude of the GW waveform. The light shaded areas show 98\% of the data, and the dark shaded areas show the interquartile range.}
 \label{strong_precession}
 \end{figure*}

During the GR precession phase of  the evolution, there are five main geometrical quantities that can affect the modulation of the emitted waveform: the precession amplitude
 $\theta_L$, precession frequency $\Omega$, the nutation amplitude $\Delta\theta_L$, the nutation frequency $\omega$, and the oscillation of the precession frequency due to nutation $\Delta\Omega$ (for details on how these quantities are defined, see \citealt{2015PhRvL.114h1103K,2015PhRvD..92f4016G,2019CQGra..36j5003G,2017PhRvD..96b4007Z}). Figure \ref{strong_precession}  shows the evolution of these quantities for the merging MBH binary population
 as a function of binary separation. 

The top left panel in Figure \ref{strong_precession} shows the evolution of the precession amplitude ($\theta_{\rm L}$) during binary inspiral. Binaries in Figure \ref{strong_precession} indicate all the merged binaries
, as we do not explicitly calculate an event rate for mergers in LISA band.
At large separations \mbox{($a\gtrsim10^4 M$)}, the median precession amplitude is small, \mbox{$\theta_{\rm L}\lesssim 10^{-3}$} rad. But $\theta_{\rm L}$ generally increases as the binary inspiral progresses, with median values of a few $\times 10^{-2}$ to 0.1 rad at $a<10^3 M$. In addition, a growing tail of large precession amplitudes appears at small separations. About 14\% of all merging binaries have a maximum $\theta_L$ greater than  $\pi/6$, indicating that a small but significant fraction of all merging MBH binaries undergo strong precession. Additionally, $\sim 5\%$ have very high maximum precession amplitudes of $\theta_{\rm L} > \pi/2$. 

All of these strongly precessing systems have misaligned spins at the onset of the GW-driven phase, originating from the isotropic distribution. 
In fact, 70\% of all such binaries with misaligned spins have a maximum $\theta_{\rm L} > \pi/6$, and 26\% of the misaligned population has a maximum $\theta_{\rm L} > \pi/2$. The median $\theta_L$ before merger for the misaligned population is 0.96 radians, while for the aligned population it is 0.04 radians. 
Note that $\theta_L$ increases as the binary inspiral progresses; thus, the maximum precession amplitude generally occurs at separations near $10 M$. 

The median nutation amplitude ($\Delta\theta_{\rm L}$; top right panel in Figure \ref{strong_precession}) similarly increases with decreasing binary separation, with typical values of a few $\times 10^{-4}$ rad at $r\gtrsim 1000 M$ to $\sim 10^{-3} - 10^{-2}$ rad at $r \sim 10 - 100 M$. The precession and nutation frequencies ($\Omega$ and $\omega$, middle row panels in Figure \ref{strong_precession}) and  $|\Delta\Omega|$ (3rd row panel shown in grey) all increase sharply with decreasing binary separation. Note that $\Delta\Omega$ can be either positive or negative, but its absolute value  increases with decreasing binary separation.  

A significant fraction of binaries experience strong precession, even in our conservative fiducial model where most binaries have aligned spins prior to merger.  This suggests that some binaries could be strongly precessing when they enter the LISA waveband. Our findings therefore provide strong motivation for future work to quantify the detectability of precession signatures in LISA waveforms.  

\section{Discussion}
\label{secdisc}

We study the evolution of spinning MBH binaries in a cosmological 
framework, considering both gas-driven spin alignment in CBDs and
relativistic precession in the GW-dominated regime. 
The initial conditions for these calculations are the MBH binary masses, separations, accretion rates, and host galaxy properties of merging MBHs from the Illustris cosmological hydrodynamics simulation  \citep[e.g.,][]{2014MNRAS.444.1518V, 2014Natur.509..177V, 2014MNRAS.445..175G, 2015A&C....13...12N}. 
MBH binary hardening rates due to dynamical friction, stellar loss-cone scattering, gaseous CBDs, 
and GWs are modeled as in \citet{2017MNRAS.464.3131K,2017MNRAS.471.4508K}. Spin evolution in the gas-driven inspiral phase is based on the Bardeen-Peterson alignment timescale and the gas properties of the host galaxy. 
Finally, we model spin precession in the GW dominated phase using a PN scheme \citep{2016PhRvD..93l4066G} and calculate the GW recoil velocity of the merged MBH.

Our key results can be enumerated as follows:

\begin{itemize}

 \item The fraction of misaligned binaries is non-negligible even in our 
 conservative fiducial model. The misaligned primaries and secondaries represent {19\%} and {10\%}, 
 respectively, of the MBH binaries that merge by $z=0$.
 This fraction is up to {$\sim$80\%} for less conservative models 
 with differing assumptions for accretion rate and disc geometry. 
 Thus, gas-driven MBH spin alignment depends strongly on the detailed conditions in the CBD. 
d The spin distribution before merger, and hence the shape of the recoil velocity distribution, is mostly determined by the disc phase of the inspiral.

 \item 
 The GW phase strongly modifies the $\Delta\Phi$ distribution, which affects recoil velocities. However, the effect on the overall recoil distribution is small, owing to the low percentage of misaligned spins in our fiducial model.
 Nonetheless, a non-negligible fraction of merging MBHs obtain large recoil velocities. 
 In our conservative fiducial model, more than {12\%} 
 of merger remnants have recoil velocities \mbox{$>500$ \kms}, 
 and more than 3\% have velocities \mbox{$>1000$ \kms}.
 This is higher than the escape velocity of most massive
  galaxies. In our least conservative model, 
 {31\%} and {14\%} of recoils have velocities \mbox{$>500$ \kms} and \mbox{$>1000$ \kms}, respectively.

 \item Strongly precessing systems constitute a significant number of binaries. We find that 14\% and 5\% of all merging binaries have a maximum precession amplitude \mbox{$\theta_L > \pi/6$} and \mbox{$\theta_L > \pi/2$} radians, respectively. In fact, the large majority (70\%) of binaries with misaligned spins at the onset of the GW phase have a maximum \mbox{$\theta_L > \pi/6$}, and 26\% of misaligned binaries have maximum \mbox{$\theta_L > \pi/2$}. Although we cannot comment directly on the potential detectability of these precessing GW waveforms with LISA, our results strongly motivate future work to quantify the likelihood that such signatures will be observable in the LISA data stream. 
 
  \item The MBH merger rate from our model is 0.15 mergers per year.
  Because we are not probing masses $< 10^6$ \msun, the actual LISA
  detection rate will be higher. Our results are in good agreement with
  similar recent analysis \citep[e.g.][]{2020MNRAS.491.2301K}.

\end{itemize}


Our findings show that there are a significant number of systems with recoil velocities higher than \mbox{500 \kms}---larger than the escape velocity of some galaxies. This indicates that MBHs may often be displaced from their host nuclei at least briefly following a merger, and it implies the existence of an intergalactic population of MBHs with no host galaxy \citep[e.g.,][]{2003ApJ...582..559V, 2004ApJ...604..484M, 2011MNRAS.412.2154B, 2015MNRAS.446...38G,2020MNRAS.495.4681I}; some of these could be observable as offset AGN \citep[e.g.,][]{2007PhRvL..99d1103L, 2008ApJ...687L..57V, 2016MNRAS.456..961B}. 
Ejected and displaced MBHs could also deflate the subsequent MBH merger rate \citep{2020arXiv200603065B}. 
\cite{2010MNRAS.404.2143V} showed that the possibility of ejection is strongly suppressed in gas rich environments where the spins are more aligned. However, in their study, they do not take into account the general relativistic evolution of spins.
Recoil velocities of merged MBHs depend strongly on the spin configurations of the progenitors.
We find that, although general relativistic spin precession can strongly affect individual binary spins, it has minimal effect on the overall recoil distribution of merging MBHs.
The main factor responsible for the changes in recoil velocities is the BP alignment in the disc-dominated phase.

The efficiency of BP alignment depends strongly 
on accretion rates and disc aspect ratios.
However, in reality these two parameters are also correlated with each other; geometrically-thin, radiatively-efficient accretion discs are commonly associated with high accretion rates \citep[e.g.,][]{1988ApJ...332..646A}. 
 Because we treat these disc parameters independently and assume that the BP effect (Equation \ref{alignment_time_scale}) can be applied to all binaries, it is possible that our model overestimates the role of BP alignment in the CBD-driven phase.

Another potential limitation of our model lies in the 
implicit assumption that, on average, the MBHs are spun up in gas rich hosts. This might not always be case, for example when MBH accretion is dominated by chaotic accretion episodes \citep[e.g.][]{2006MNRAS.373L..90K, 2008ApJ...684..822B, 2011MNRAS.410...53F} or irregular flows caused by angular momentum flips during galaxy mergers \citep{2017MNRAS.465.2643C}. 
However, because spin orientations evolve on much shorter timescales than spin magnitudes, the coherence of larger-scale accretion flows  
is likely to affect the spin magnitudes more than the spin orientations.
We recall that our results depend very minimally on the choice of spin
magnitudes.
Our assignments of the spin magnitudes could also be improved by considering a model 
in which the spin evolution due to accretion is explicitly traced through the CBD phase.
We refer the reader to  \citet{2014MNRAS.440.1590D} and \citet{2019MNRAS.490.4133B} for a more in depth discussion of accretion and merger effects on the spins. In addition to that we have also not considered the case of anti-alignment of the disk and MBH. Depending on the mass of the MBH and the disk mass the accretion could be episodic and the disk might align or anti-align with MBH. This can lead to either spin-up or spin-down of the MBH \citep{2018MNRAS.477.3807F}. 


In the GW dominated phase we use a PN scheme that does not evolve the binary eccentricities; the analytic calculation of eccentricity evolution is done separately for the GW phase.  This is a reasonable approximation as GW tend to circularize binaries on a timescale which is shorter than the inspiral time \citep[][see also Fig.~\ref{fig:eccentricity_evolution}]{1964PhRv..136.1224P}. Additionally, \citet{2019PhRvD.100l4008P} have recently shown that eccentricity is subdominant in the spin morphology evolution of MBH binaries. We hope to include a treatment of spinning eccentric binaries in future work. 

When the MBH binary inspiral time is longer than the typical time between galaxy mergers, a triple MBH system may form.
\citet{2017MNRAS.464.3131K} find that a non-negligible fraction of binaries are still unmerged when a subsequent galaxy merger occurs, but as in that work, we do not attempt to model triple MBH systems here. Triples may not only affect eccentricities but also have important consequences for merger rates. In a triple system, the lightest MBH can get ejected out of the system and accelerate the shrinking of the binary separation \citep{1975AJ.....80..809H}. 
Alternatively, a third MBH can settle into an outer semi-circular orbit and form a hierarchical configuration. 
The outer MBH can then accelerate the hardening of the inner binary \citep{1962P&SS....9..719L, 1962AJ.....67..591K, 2002ApJ...578..775B}. These factors can increase the overall merger rates \citep[e.g.][]{2019MNRAS.486.4044B,2019MNRAS.487.4985B}. 
Kozai-Lidov oscillations between eccentricity and inclination of the inner binary can also lead to large spin misalignments  \citep{2018ApJ...863....7R, 2018ApJ...863...68L, 2019ApJ...881...41L}.

In summary, our results demonstrate that MBH spins are a crucial aspect of MBH binary evolution, which will impact the observability of MBH binaries as GW and multi-messenger sources for LISA. We find that misaligned spins are not a rare occurrence over cosmic time, suggesting that large recoil velocities may reduce the MBH merger rate somewhat and produce a population of offset or wandering MBHs. Some of these may be observable as offset AGN. The misaligned binary population in our models also suggests that some binaries may be strongly precessing in the LISA band, which could potentially be detected in their GW waveforms. Any such detections would place strong constraints on MBH spins and provide direct confirmation of GR precession. Precessing, accreting binaries could also produce unique electromagnetic signatures such as precessing jets \citep[e.g.,][]{1982ApJ...262..478G, 2019MNRAS.482..240K} or the shape and variability of Fe K$\alpha$ profiles \citep{2001A&A...377...17Y}. Future work to refine and quantify these predictions in advance of LISA will therefore provide key information about the GW event rate and source characteristics.

\section*{Acknowledgements}

We would like to thank 
the anonymous referee for their helpful suggestions that have improved the quality of this manuscript. We would also like to thank
Pedro Capello, Chiara Mingarelli, Dan D'Orazio
, and Marta Volonteri
for insightful comments and discussions. 
In addition, we would like to thank the attendees of the LISA Symposium (2018), BASS workshop (2019), and JSI workshop (2019) for fruitful suggestions. 
This work made use of the \textsc{python} programming language \citep{python}, and its  
\textsc{numpy} \citep{2011CSE....13b..22V}, \textsc{scipy} \citep{2020NatMe..17..261V} and \textsc{Matplotlib} \citep{2007CSE.....9...90H} packages along with
Jupyter notebooks \citep{soton403913}.
L.B. acknowledges support from NSF Grant No. AST-1909933.
D.G. is supported by European Union's H2020  ERC Starting Grant No. 945155--GWmining, Leverhulme Trust Grant No. RPG-2019-350, and Royal Society Grant No. RGS-R2-202004.
M.K. is supported by NSF Grants No. PHY-1607031 and PHY-2011977.
Computational work was 
performed on the University of Florida Hipergator cluster,
Harvard's Odyssey cluster, 
the University of Brimingham BlueBEAR cluster, the Athena cluster at HPC Midlands+ funded by EPSRC Grant No. EP/P020232/1, and the Maryland Advanced Research Computing Center (MARCC).

\section*{Data Availability}
The data underlying this article will be shared on reasonable request to the corresponding author.

\bibliographystyle{mnras_tex_edited}
\bibliography{main} %

\label{lastpage}
\end{document}